\documentclass[lettersize,journal]{IEEEtran}
\usepackage{stmaryrd}
\usepackage{lipsum}  
\usepackage{svg}
\usepackage{amssymb,amsmath,amsfonts,wasysym,latexsym}
\usepackage[acronyms,shortcuts]{glossaries}
\usepackage{bm}
\usepackage{tikz}
\usetikzlibrary{patterns,arrows,decorations.pathreplacing}
\usepackage{psfrag}
\usepackage{amssymb}
\newcommand{\argmin}{\mathop{\rm argmin}}
\newcommand{\normof}[2]{\left\|#1\right\|_{#2}}
\newcommand{\fronorm}[1]{\normof{#1}{\rm F}}            
\newcommand{\blockron}{\hspace{0.05cm}|\hspace{-0.1cm}\otimes\hspace{-0.1cm}| \hspace{0.05cm}}
\usepackage{epstopdf}
\usepackage{amsmath}
\usepackage{scalefnt}
\usepackage{color}
\usepackage{xcolor}
\usepackage{algorithm}
\usepackage{algpseudocode}
\usepackage{cite}
\usepackage{caption}
\usepackage{subcaption}
\usepackage{caption}
\usepackage{multirow}
\usepackage{graphicx}

\definecolor{forestgreen}{rgb}{0, 0.5,0.5} 
\definecolor{darkgreen}{rgb}{0,0.392157,0} 

\newcommand{\nmode}[2]{\big[\ma{\mathcal{#1}}\big]_{\left(#2\right)}}
\newcommand{\sumx}[2]{\sum\limits_{#1}^{#2}}
\newcommand{\bb}[1]{\mathbb{#1}}
\newcommand{\ten}[1]{\boldsymbol{\mathcal #1}}
\newcommand{\ma}[1]{\boldsymbol{#1}}
\definecolor{green}{rgb}{0.1,0.75,0.2}

\newcommand{\unvec}[1]{\text{unvec}\big( #1 \big)}
\renewcommand{\vec}[1]{\text{vec}\big( #1\big)}

\newacronym{2G}{2G}{second generation}
\newacronym{3G}{3G}{third generation}
\newacronym{4G}{4G}{fourth generation}
\newacronym{5G}{5G}{fifth generation}
\newacronym{B5G}{B5G}{beyond fifth generation}
\newacronym{6G}{6G}{sixth generation}
\newacronym{3GPP}{3GPP}{3$\text{rd}$~Generation Partnership Project}
\newacronym{LTE}{LTE}{long term evolution}
\newacronym{NR}{NR}{new radio}
\newacronym{LS}{LS}{least squares}

\newacronym{IRS}{IRS}{intelligent reconfigurable surface}
\newacronym{RIS}{RIS}{reconfigurable intelligent surface}
\newacronym{LIS}{LIS}{large intelligent surface}
\newacronym{SDS}{SDS}{software-defined surface}

\newacronym{D2D}{D2D}{device-to-device}
\newacronym{BS}{BS}{base station}
\newacronym{UE}{UE}{user equipment}

\newacronym{SU}{SU}{single-user}
\newacronym{MU}{MU}{multi-user}
\newacronym{SISO}{SISO}{single-input single-output}
\newacronym{MISO}{MISO}{multiple-input single-output}
\newacronym{SIMO}{SIMO}{single-input multiple-output}
\newacronym{MIMO}{MIMO}{multiple-input multiple-output}

\newacronym{CSI}{CSI}{channel state information}
\newacronym{LOS}{LOS}{line-of-sight}
\newacronym{NLOS}{NLOS}{non-line-of-sight}

\newacronym{QoS}{QoS}{quality-of-service}
\newacronym{SE}{SE}{spectral efficiency}
\newacronym{EE}{EE}{energy efficiency}
\newacronym{SINR}{SINR}{signal to interference plus noise ratio}
\newacronym{SNR}{SNR}{signal to noise ratio}

\newacronym{ProSe}{ProSe}{proximity services}
\newacronym{NSPS}{NSPS}{national security and public safety}

\newacronym{RRM}{RRM}{radio resource management}
\newacronym{MS}{MS}{mode selection}
\newacronym{RA}{RA}{resource allocation}
\newacronym{PC}{PC}{power control}

\newacronym{BCD}{BCD}{block coordinate descent}

\newacronym{RF}{RF}{radio frequency}
\newacronym{AWGN}{AWGN}{additive white Gaussian noise}
\newacronym{MRC}{MRC}{maximum ratio combining}

\newacronym{AF}{AF}{amplify-and-forward}
\newacronym{DF}{DF}{decode-and-forward}
\newacronym{DFT}{DFT}{discrete Fourier transform}
\newacronym{TX}{TX}{transmitter}
\newacronym{RX}{RX}{receiver}
\newacronym{ALS}{ALS}{alternating least squares}
\newacronym{BALS}{BALS}{bilinear alternating least squares}
\newacronym{SVD}{SVD}{singular value decomposition}
\newacronym{HOSVD}{HOSVD}{high order singular value decomposition}
\newacronym{THOSVD}{THOSVD}{truncated high order singular value decomposition}
\newacronym{PARAFAC}{PARAFAC}{PARAllel FACtors}
\newacronym{AOD}{AOD}{angle of departure}
\newacronym{AOA}{AOA}{angle of arrival}
\newacronym{URA}{URA}{uniform rectangular array} 
\newacronym{ADR}{ADR}{achievable data rate}
\newacronym{NMSE}{NMSE}{normalized mean square error}
\newacronym{SER}{SER}{symbol error rate}
\newacronym{LRA}{LRA}{low-rank approximation}

\newacronym{ULA}{ULA}{uniform linear array}
\newacronym{mmWave}{mmWave}{milimiter-wave}
\newacronym{CS}{CS}{compressed sensing}
\newacronym{OFDM}{OFDM}{orthogonal frequency division multiplexing}
\newacronym{PIN}{PIN}{positive-intrinsic-negative}
\newacronym{BD-RIS}{BD-RIS}{beyond diagonal reconfigurable intelligent surface}
\newacronym{LS-Kron}{LS-Kron}{least squares Kronecker factorization}

\newacronym{BTALS}{BTALS}{block Tucker alternating least squares}
\newacronym{BTKF}{BTKF}{block Tucker Kronecker factorization}
\newacronym{PALS}{PALS}{PARAFAC alternating least squares}
\newacronym{PKF}{PKF}{PARAFAC Khatri-Rao factorization}
\newacronym{ISAC}{ISAC}{integrated sensing and communication}

\usepackage[none]{hyphenat}
\begin{document}
\title{ 
Channel Estimation for Beyond Diagonal RIS\\via Tensor Decomposition
 } 

\author{André L. F. de Almeida,~\IEEEmembership{Senior Member,~IEEE,}  Bruno Sokal,~\IEEEmembership{Member,~IEEE,}\\
Hongyu Li,~\IEEEmembership{Graduate Student Member,~IEEE,} and Bruno Clerckx,~\IEEEmembership{Fellow,~IEEE}
\thanks{This work was partially supported by Fundação Cearense de Apoio ao Desenvolvimento Científico e Tecnológico (FUNCAP) under grants INCT-25255-82587.32.41/64, ITR-0214-00041.01.00/23, and the National Institute of Science and Technology (INCT-Signals) sponsored by Brazil's National Council for Scientific and Technological Development (CNPq) under grant 406517/2022-3. This work is also partially supported by CNPq under grants 312491/2020-4 and 443272/2023-9.
 This work has been partially supported by the UKRI under Grant EP/Y004086/1, EP/X040569/1, EP/Y037197/1,EP/X04047X/1, EP/Y037243/1.}}

\renewcommand{\arraystretch}{1.2} 

\maketitle

\begin{abstract}
This paper addresses the channel estimation problem for beyond diagonal reconfigurable intelligent surface (BD-RIS) from a tensor decomposition perspective. We first show that the received pilot signals can be arranged as a three-way tensor, allowing us to recast the cascaded channel estimation problem as a block Tucker decomposition problem that yields decoupled estimates for the involved channel matrices while offering a substantial performance gain over the conventional (matrix-based) least squares (LS) estimation method. More specifically, we develop two solutions to solve the problem. The first is a closed-form solution that extracts the channel estimates via a block Tucker Kronecker factorization (BTKF), which boils down to solving a set of parallel rank-one matrix approximation problems. Exploiting such a low-rank property yields a noise rejection gain compared to the standard LS estimation scheme, allowing the two involved channels to be estimated separately. The second solution is based on a block Tucker alternating least squares (BTALS) algorithm that directly estimates the involved channel matrices using an iterative estimation procedure. We discuss the uniqueness and identifiability issues and their implications for training design. We also propose a tensor-based design of the BD-RIS training tensor for each algorithm that ensures unique decoupled channel estimates under trivial scaling ambiguities. Our numerical results shed light on the tradeoffs offered by BTKF and BTALS methods. Specifically, while the first enjoys fast and parallel extraction of the channel estimates in closed form, the second has a more flexible training design, allowing for a significantly reduced training overhead compared to the state-of-the-art LS method.
\end{abstract}

\begin{IEEEkeywords}
Beyond diagonal reconfigurable intelligent surfaces, channel estimation, tensor decomposition, alternating least squares, Kronecker factorization.
\end{IEEEkeywords}

\renewcommand\baselinestretch{.85}

\section{Introduction}\label{Sec:Introduction}
As a new advance of conventional reconfigurable intelligent surface (RIS) techniques with diagonal phase shift matrices \cite{basar2019,jian2022reconfigurable}, beyond diagonal (BD) RIS has been recently proposed and theoretically proved to achieve enhanced channel gain and enlarged coverage \cite{Clerckx_CM_MAR_2024,CLX_BD_GC_2023,Clerckx_Arxiv_2024}.
The benefits of BD-RIS are enabled by interconnecting elements via additional tunable components to mathematically generate scattering matrices with nonzero off-diagonal entries, increasing flexibility to manipulate waves. The fundamental modeling and architecture (group/fully connected) design of BD-RIS based on the circuit topology has been first studied in \cite{shen2021modeling}. Following \cite{shen2021modeling}, other architectures (forest/tree-connected) with reduced circuit complexity yet satisfactory performance have been proposed based on the graph theory \cite{Clerckx_TWC_EA_2024}. \textcolor{black}{In \cite{de2024beyond}, a frequency-dependent model is proposed to optimize a BD-RIS architecture for multi-band MIMO networks.}
Meanwhile, BD-RIS with hybrid transmissive, reflective, and multi-sector modes have been proposed to enlarge the coverage, based on the antenna array arrangement \cite{Clerckx_CM_MAR_2024}.

It is worth noting that the enhanced performance of BD-RIS architectures and modes depends highly on the accuracy of the channel state information (CSI). However, it is difficult to effectively and efficiently acquire the CSI for BD-RIS-aided wireless systems for the following reasons. 
\textit{First}, since the BD-RIS is nearly passive without the ability to sense signals, a straightforward strategy is to estimate the \textit{composite} channel constructed by the transmitter-RIS and RIS-user channels, as well studied in conventional RIS literature \cite{swindlehurst2022channel,zheng2022survey,RIS_sigpro_Lee_2022}, \cite{C_HU_2019,Alexandropoulos_2020}. This strategy relies on the pre-design of BD-RIS for pilot training. At the same time, each BD-RIS architecture leads to unique mathematical constraints of the scattering matrix, which indicates that the design for conventional RISs does not work for BD-RIS architectures.  
\textit{Second}, each BD-RIS architecture leads to unique constructions of the composite channel with increasing dimensions, which requires additional training overhead to obtain the CSI.
To solve the above two challenges, one recent work \cite{Clerckx_Arxiv_2024} has proposed a closed-form solution based on the least squares (LS) estimation to pre-design the BD-RIS with group/fully-connected architectures.
Nevertheless, there are two limitations in \cite{Clerckx_Arxiv_2024}. 
\textit{First}, the composite CSI is obtained at the cost of a large training overhead, which grows heavily with the circuit complexity of BD-RIS architectures.
\textit{Second}, the built-in block Kronecker structure of the composite channel is ignored, which could be further exploited to facilitate channel estimation.
Thus, estimating the BD-RIS-aided channels with high accuracy and low training overhead remains an essential yet challenging open problem.

To address the above limitations, this paper studies channel estimation for BD-RIS via a tensor decomposition perspective.

Tensor modeling has been widely adopted in signal processing for wireless communications due to its ability to manipulate and efficiently exploit the intrinsic multidimensional structure of wireless communication signals and channels (usual dimensions are space, time, frequency, polarization, coding, etc.) \cite{Sidiropoulos_DSCDMA,Almeida_Elsevier_2007,confac,favier2014tensor,nestedparafac,FavierAlmeida2014,Sidiropoulos2017,Miron2020,Chen2021}. In the case of \ac{RIS}, tensor decompositions have proven to provide effective solutions to channel estimation and beamforming design \cite{gil2021,alexandropoulos2020hardware,Araujo2023,Sokal_Overhead,benicio2023tensor_wcl}. This is due to the ability of tensors to decouple individual estimates of the involved channels using decompositions such as the \ac{PARAFAC} and Tucker decompositions under more flexible conditions than competing methods.
{\color{black}Tensor-based algorithms have recently been used in \ac{ISAC} systems \cite{R2_ISAC_TB,R3_ISAC_Terahetz,R4_DMA}. The work of \cite{R2_ISAC_TB} models the received sensing signal using a PARAFAC decomposition, and the authors propose an \ac{ALS} algorithm to estimate the sensing factors. On the other hand, in \cite{R3_ISAC_Terahetz} the authors exploit the sparsity of the THz channels and model the signal using a PARAFAC tensor while proposing an \ac{ALS} based algorithm to estimate the channel parameters. In \cite{R4_DMA}, a PARAFAC tensor model is proposed for a channel estimation framework in MIMO-OFDM with dynamic metasurface antennas (DMA). Different from \cite{R2_ISAC_TB,R3_ISAC_Terahetz,R4_DMA}, the proposed approach focuses on channel estimation for \ac{BD-RIS} assisted \ac{MIMO} systems, by employing a tensor-based model which allows for a decoupled estimate of the involved channels.}

{\color{black}It is worth noting that in the case of single-connected (diagonal) RIS, the solution proposed in \cite{gil2021} showed that the individual channel estimation problem could be solved using the \ac{PARAFAC} decomposition, allowing us to find decoupled channel estimates of the involved channels, while providing a refined estimate of the composite TX-RIS-RX channel. However, individual channel estimation methods valid for traditional single-connected RIS do not apply to BD-RIS due to its more general (and complex) inter-element connection structure and design constraints. More specifically, the PARAFAC-based individual channel estimation method of \cite{gil2021} cannot capture the algebraic tensor structure of the received pilots, which requires rethinking channel training in two aspects. First, an adequate tensor decomposition approach is necessary to obtain decoupled channel estimates from the received pilot signals, and second, BD-RIS training structures are equivalent to third-order tensors (as opposed to diagonal RIS training structures modeled as matrices).

This paper shows that the composite channel for BD-RIS can be formulated as a (block) Tucker tensor decomposition. Decoupled estimates for the involved channel matrices can be obtained by exploiting the different unfoldings of the received pilot tensor. We develop two solutions to solve this problem. The first is a closed-form solution that extracts the individual channel estimates via a \ac{BTKF}, which boils down to solving a set of parallel rank-one matrix approximation problems. The second one is based on a \ac{BTALS} algorithm that directly estimates the involved channels using an iterative estimation procedure. The proposed algorithms offer substantial performance gains over conventional LS estimation while operating at much lower training overheads by capitalizing on the tensor signal structure.}

The contributions of this paper are summarized as follows:

\textit{First}, we link the channel estimation problem for BD-RIS to a tensor decomposition problem. Specifically, we show that the received pilot signals can be organized as a three-dimensional (3D) array or a third-order tensor that follows a (block) Tucker decomposition model. In addition, we discuss the implications of the specific BD-RIS architecture on the resulting tensor decomposition structure.

\textit{Second}, we propose two tensor decomposition-based channel estimation schemes for BD-RIS that capitalize on tensor modeling. The  \ac{BTKF} method is a closed-form scheme that yields decoupled estimates of the involved channels after an LS estimation step by solving a block-Kronecker factorization problem that boils down to solving rank-one approximation problems based on the $3$-mode unfolding of the received pilot tensor. The second method, \ac{BTALS}, directly estimates the channel matrices separately using the 1-mode and 2-mode unfoldings of the received pilot tensor. 

\textit{Third}, we discuss the trade-offs involving the channel estimation methods and their implications for the training design.
On the one hand, we show that the BTKF method effectively exploits the inherent Kronecker structure of the composite channel to provide a more accurate reconstruction than the reference LS method, thanks to the noise rejection property of the channel separation step.
On the other hand, by efficiently exploiting the tensor decomposition structure of the received pilots, BTALS yields decoupled channel estimates with a much lower training overhead, which can be orders of magnitude smaller than that of LS and BTKF methods. Such savings in training resources offered by BTALS are even more pronounced for BD-RIS configurations with higher levels of couplings among the scattering elements. 

\textit{Fourth}, we study the identifiability conditions and uniqueness associated with the proposed algorithms and their implications for training design. We propose a new training design for the BD-RIS scattering matrix and derive the structure of the BD-RIS training tensor used in each proposed channel estimation method. The proposed designs fulfill the physical constraints of the BD-RIS architecture and the identifiability conditions of the associated block Tucker model.

\textit{Finally}, simulation results show the superiority of the proposed \ac{BTKF} and \ac{BTALS} algorithms compared to the baseline LS estimator while highlighting the trade-offs involving these methods in terms of \ac{NMSE} performance, required training overhead, and computational complexity. In particular, we show that BTKF and BTALS offer performance gains over the reference LS method thanks to decoupling the estimated channel matrices, yielding a more accurate reconstruction of the composite channel. For a group-connected BD-RIS architecture with $Q$ groups, the BTKF method achieves channel separation by solving a set of $Q$ rank-one matrix approximations in parallel. On the other hand, the BTALS method directly estimates the involved channel matrices by intertwining the estimation of the transmitter-RIS and RIS-receiver channels in an iterative way using an alternating least squares procedure. Our results also show that the NMSE performance of the estimated channels with group-connected BD-RIS architectures is the same as that with conventional RIS in some scenarios.

This work is organized as follows. Section II provides the basic material and definitions related to tensor decomposition, along with the main notations and properties. Section III describes the system model and discusses the baseline LS method and BD-RIS design. Section IV gives a detailed presentation of the tensor modeling of the received pilot signals. Section V formulates the proposed channel estimation methods. This section also discusses computational complexity and uniqueness issues. Section VI describes the proposed BD-RIS tensor design. Numerical results are discussed in Section VII, and the paper is concluded in Section VIII.

\section{Tensor Prerequisites}\label{Sec:Tensor_background}
In this section, we provide the proper notations and main operators used in this paper and an overview of the Tucker decomposition, which will be used to develop the proposed channel estimation algorithm.

\subsection{Notation and properties} \label{Sec:notation}
Scalars are represented as non-bold lower-case letters $a$, column vectors as lower-case boldface letters $\ma{a}$, matrices as upper-case boldface letters $\ma{A}$, and tensors as calligraphic upper-case letters $\ten{A}$. The superscripts $\{\cdot\}^{\text{T}}$, $\{\cdot\}^{\text{*}}$, $\{\cdot\}^{\text{H}}$ and $\{\cdot\}^{+ }$ stand for transpose, conjugate, conjugate transpose, and pseudo-inverse operations, respectively. \textcolor{black}{An identity matrix of dimension $K$ is denoted as $\ma{I}_{K}$.} The operator $\Arrowvert\cdot\Arrowvert_{\text{F}}$ denotes the Frobenius norm of a matrix or tensor, and $\bb{E}\{\cdot\}$ is the expectation operator.
\textcolor{black}{Given a matrix $\ma{A} \in \bb{C}^{I \times R}$, the operator $\text{D}_{i}(\ma{A})$ defines a diagonal matrix of size $R \times R$ constructed from the $i$-th row of $\ma{A}$, for $i \in \{1,\ldots,I\}$.} From a set of $Q$ matrices $\ma{X}^{(q)} \in \bb{C}^{M \times N}$, $q = \{1,\ldots,Q\}$, we can construct a block diagonal matrix as \textcolor{black}{ $\ma{X} = \text{blkdiag}\big(\ma{X}^{(1)},\ldots,\ma{X}^{(Q)}\big)  \in \bb{C}^{MQ \times NQ}$}. Moreover, $\text{vec}\left(\ma{A}\right)$ converts $\ma{A} \in \mathbb{C}^{I \times R}$ to a column vector $\ma{a} \in \mathbb{C}^{IR \times 1}$ by stacking its columns on top of each other, while the operator \textcolor{black}{$\unvec{\ma{a}}_{I \times R}$} returns to the matrix $\ma{A} \in \bb{C}^{I \times R}$. The symbols $\circ$, $\otimes$,  $\diamond$, and $\blockron$ denote the outer product, the Kronecker product, the Khatri-Rao product (also known as the column-wise Kronecker product), and the block Kronecker product, respectively. The Khatri-Rao product of matrices $\ma{X} \in \bb{C}^{I \times R}$ and $\ma{Y} \in \bb{C}^{J \times R}$, is defined as $\ma{Z} = \ma{X} \diamond \ma{Y} = [\ma{x}_1 \otimes \ma{y}_1, \ldots,\ma{x}_R  \otimes \ma{y}_R] \in \bb{C}^{JI \times R}$,
where $\ma{x}_r$ and $\ma{y}_r$ are the $r$-th column of $\ma{X}$ and $\ma{Y}$, respectively, $r=1,\ldots,R$. Likewise, let us define  $\ma{H} = [\ma{H}^{(1)},\ldots, \ma{H}^{(Q)}] \in \bb{C}^{M \times LQ}$ and \textcolor{black}{$\ma{G}= [\ma{G}^{(1)},\ldots, \ma{G}^{(Q)}] \in \bb{C}^{N \times LQ}$,} matrices formed each by $Q$ block matrices, i.e., $\ma{H}^{(q)} \in \bb{C}^{M \times L}$  and $\ma{G}^{(q)} \in \bb{C}^{N \times L}$, $q= 1,\ldots,Q$. The block Kronecker product\footnote{The block Kronecker product defined in (\ref{eq:blkron}) is also referred to in the literature as the Khatri-Rao product between partitioned matrices \cite{Lathauwer18blockII}.} between $\ma{H}$ and $\ma{G}$, denoted as $\ma{W} = \hspace{-0.1cm}\ma{H} \blockron \ma{G}$, is given by
\begin{equation}\label{eq:blkron}
    \ma{W} =  [\ma{H}^{(1)} \otimes \ma{G}^{(1)},\ldots,\ma{H}^{(Q)} \otimes \ma{G}^{(Q)}] \in \bb{C}^{NM \times L^2Q}.
\end{equation}
We also use the following property of the Kronecker product
\begin{align}
\label{prop:vec}\text{vec}\left(\ma{A}\ma{B}\ma{C}\right) &= \left(\ma{C}^{\text{T}} \otimes \ma{A}\right)\text{vec}\left(\ma{B}\right),
\end{align}
where the involved matrices have compatible dimensions.

\subsection{Slices and unfoldings}
Consider a set of matrices $\ma{Y}_{k} \in \bb{C}^{I \times J}$, \textcolor{black}{$\forall k = 1,\ldots, K$.} Concatenating all $K$ matrices, we form the third-order tensor $\ten{Y} = \ma{Y}_1 \sqcup_3 \ma{Y}_2 \sqcup_3 \ldots \sqcup_3 \ma{Y}_{K} \in \bb{C}^{I \times J \times K}$, where $\sqcup_3 $ indicates a concatenation along the third dimension. We can interpret $\ma{Y}_{k}$ as the $k$-th frontal slice of $\ten{Y}$, defined as the matrix $\ten{Y}_{..k} = \ma{Y}_{k} \in \bb{C}^{I \times J}$. This matrix is built by varying the first and second dimensions for a fixed third-dimension index $k$. The tensor $\ten{Y}$ can be \textit{matricized} by letting one dimension vary along the rows and the remaining two dimensions along the columns. From $\ten{Y}$, we can form three different matrices, referred to as the \textit{$n$-mode unfolding}, $n=1,2,3$, which can be respectively obtained as a function of the frontal slices as
\begin{align}
\label{eq:nmode_1}\nmode{Y}{1} &= [\ten{Y}_{..1},\ldots,\ten{Y}_{..K}] \in \bb{C}^{I \times JK}, \\  
\label{eq:nmode_2}\nmode{Y}{2} &= [\ten{Y}_{..1}^{\text{T}},\ldots,\ten{Y}_{..K}^{\text{T}}] \in \bb{C}^{J \times IK},\\ 
\label{eq:nmode_3}\nmode{Y}{3} &= [\text{vec}(\ten{Y}_{..1}),\ldots,\text{vec}(\ten{Y}_{..K})]^{\text{T}} \in \bb{C}^{K \times IJ}.
\end{align}

For convenience, we can also refer to the unfolding operation as $\nmode{Y}{n}=\textrm{unfold}(\ten Y,n)$, $n=1,2,3$.
The $n$-mode product, denoted as ``$\times_n$", defines the multiplication between a tensor $\ten{Y}$ and a matrix $\ma{A}$, leading to a tensor $\ten{Z}$ with compatible dimensions, i.e., $\ten{Z} = \ten{Y} \times_n \ma{A}$. It can be computed by pre-multiplying the $n$-mode unfolding of $\ten{Y}$ by the matrix $\ma{A}$, i.e.,  $[\ten{Z}]_{(n)} = \ma{A}[\ten{Y}]_{(n)}$. 
For example, the 1-mode product between $\ten{Y}\in\mathbb{C}^{I\times J\times K}$ and $\ma{A}\in\mathbb{C}^{L\times I}$ yields $\ten{Z} =\ten{Y} \times_1 \ma{A}\in\mathbb{C}^{L\times J\times K}$. It can be computed by $[\ten{Z}]_{(1)} = \ma{A}[\ten{Y}]_{(1)}\in\mathbb{C}^{L\times JK}$. We refer the interested reader to \cite{Sidiropoulos2017} for an overview.

\subsection{Tucker decomposition}\label{Sec:Tucker_decom}
\textcolor{black}{The Tucker decomposition \cite{Kolda2009} defines the concept of multilinear transformation.
For a third-order tensor $\ten{Y} \in \bb{C}^{I \times J \times K}$, it expresses the tensor as multiple sums of rank-one tensor components, which can be defined as 
 \begin{align}
\ten{Y} &= \sumx{r_1=1}{R_1} \sumx{r_2=1}{R_2} \sumx{r_3=1}{R_3}g_{r_1,r_2,r_3}\ma{a}_{r_1} \circ \ma{b}_{r_2} \circ   \ma{c}_{r_3} 
,\label{eq:tucker_model} 
\end{align}
where $\ma{a}_{r_1} \in \bb{C}^{I \times 1}$, $\ma{b}_{r_2} \in \bb{C}^{J \times 1}$, and $\ma{c}_{r_3} \in \bb{C}^{K \times 1}$ are the column vectors of the factor matrices $\ma{A} \in \bb{C}^{I \times R_1}$, $\ma{B} \in \bb{C}^{J \times R_2}$, and $\ma{C} \in \bb{C}^{K \times R_3}$, respectively, and $\ten{G} \in \bb{C}^{R_1 \times R_2 \times K}$ is referred to as the \emph{core tensor}, with typical element $g_{r_1,r_2,r_3}\doteq [\ten G]_{r_1,r_2,r_3}$.
Adopting the $n$-mode product notation, the Tucker decomposition can be written as
\begin{align}
  \label{eq:tenZ_tucker_generic} \ten{Y} &= \ten{G} \times_1 \ma{A} \times_2 \ma{B} \times_3 \ma{C} 
\end{align}
A special case is the Tucker-2 decomposition, where one of its factor matrices equals the identity matrix, e.g., $\ma{C} = \ma{I}_K \in \bb{R}^{K \times K}$ (with $K=R_3$). In this case, (\ref{eq:tenZ_tucker_generic}) simplifies to
 \begin{align}
  \label{eq:tenZ_tucker_2} \ten{Y} &= \ten{G} \times_1 \ma{A} \times_2 \ma{B} 
\end{align}
The $3$-mode (frontal) slices $\ten{Y}_{..k} \in \mathbb{C}^{I \times J}$ can be expressed as
\begin{align}
   \label{eq:tucker_2_slice} \ten{Y}_{..k} &= \ma{A}\ten{G}_{..k}\ma{B}^{\text{T}} \in \bb{C}^{I  \times J}, \,\, k=1, \ldots, K.
\end{align}
}
By properly staking these frontal slices according to equations (\ref{eq:nmode_1})-(\ref{eq:nmode_3}), the three matrix unfolding of the Tucker-2 decomposition can be factorized as
\begin{align}
     \label{eq:tucker_2_u1} \nmode{Y}{1} &= \ma{A}\nmode{G}{1}\big(\ma{I}_{K} \otimes \ma{B} \big)^{\text{T}} \in \bb{C}^{I \times JK}, \\
    \label{eq:tucker_2_u2} \nmode{Y}{2} &= \ma{B}\nmode{G}{2}\big(\ma{I}_{K} \otimes \ma{A} \big)^{\text{T}} \in \bb{C}^{J \times IK}, \\
    \label{eq:tucker_2_u3} \nmode{Y}{3} &= \nmode{G}{3}\big(\ma{B} \otimes \ma{A} \big)^{\text{T}} \in \bb{C}^{K \times IJ}.
\end{align}

The Tucker decomposition is not unique due to rotational freedom involving the factor matrices and the core tensor. Indeed, the multiplication of each factor matrix by a nonsingular matrix is compensated by transforming the corresponding modes of the core tensor by the inverse of these matrices without changing the output tensor \cite{Kolda2009}. 
However, when the core tensor $\ten{G}$ is known, a unique estimation of the factor matrices under trivial scaling ambiguities is possible under certain conditions \cite{FavierAlmeida2014}, which is the case of this work.

It is worth mentioning that the well-known parallel factor (PARAFAC) decomposition, also known as the canonical polyadic decomposition (CPD)  \cite{Harshman}, \cite{Sidiropoulos2017}, is a special case of the Tucker decomposition, in which $R_1=R_2=R_3=R$, while the core tensor reduces to an identity tensor. In this case, we have
$\ten{Y} = \ten{I}_{3,R} \times_1 \ma{A} \times_2 \ma{B} \times_3 \ma{C} \in \bb{C}^{I \times J   \times K}$, with $\ma{A} \in \mathbb{C}^{I \times R}$, $\ma{B} \in \mathbb{C}^{J \times R}$, and $\ma{C} \in \mathbb{C}^{K \times R}$ being the associated factor matrices, where $R$ denotes the tensor rank.

\section{System Model}\label{Sec:System_Model}
\begin{figure}[!t]
	\centering\includegraphics[scale=0.420]{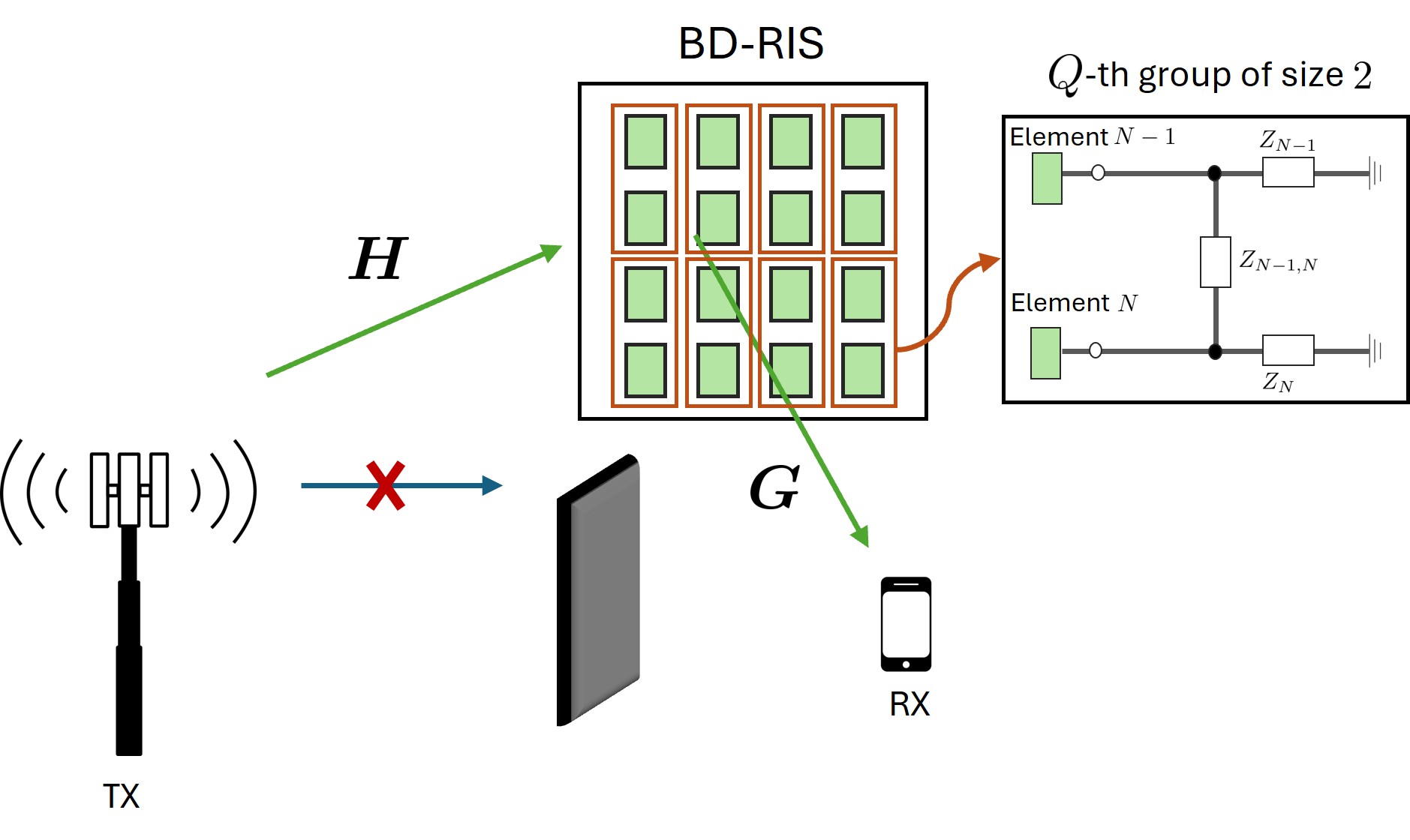}
	\caption{\small A communication system aided by a BD-RIS.}
	\label{fig:system_model}
\end{figure}

Let us consider a \ac{MIMO} system assisted by a \ac{BD-RIS}, as illustrated in Fig. \ref{fig:system_model}, where the transmitter and the receiver are equipped with $M_T$ and $M_R$ antennas, respectively, and a \ac{BD-RIS} with $N$ elements. For simplification, we assume that the direct link between the transmitter and the receiver is blocked. 

\subsection{Signal and channel models}
We adopt a two-timescale protocol, where the transmission consists of $K$ blocks of consecutive $T$ time slots each. \textcolor{black}{We assume that the length-$T$ pilot sequences are repeatedly transmitted through each block, while the BD-RIS response repeats itself within one block and varies among blocks}. This training protocol follows the idea proposed in \cite{gil2021} for diagonal RIS, \textcolor{black}{in which the scattering matrix of the BD-RIS} is fixed during the time window corresponding to a block of $T$ time slots while varying between blocks. The received pilot signal at the $t$-th time slot and $k$-th block is $\bar{\ma{y}}_{t,k} = \ma{G}\ma{S}_{k}\ma{H}^{\text{T}}\ma{x}_{t} + \bar{\ma{b}}_{t,k} \in \bb{C}^{M_R \times 1}$, \textcolor{black}{where $\ma{H} \in \bb{C}^{M_T \times N}$ and $\ma{G} \in \bb{C}^{M_R \times N}$ are the TX-RIS and RIS-RX channels, respectively, and $\ma{S}_k \in \bb{C}^{N \times N}$ is the BD-RIS scattering matrix associated with the $k$-th training block, with $\ma{S}^{\text{H}}_{k}\ma{S}_{k} = \ma{I}_{N}$.}
Collecting the $T$ time slots of the $k$-th block yields $\bar{\ma{Y}}_k=\ma{G}\ma{S}_k\ma{H}^{\text{T}}\ma{X} + \bar{\ma{B}}_k \in \bb{C}^{M_R \times T}$. 
The matrix
$\ma{X} = [\ma{x}_1, \ldots, \ma{x}_T] \in \bb{C}^{M_T \times T}$ collects the transmitted pilot symbols during the $T$ time slots, and $\bar{\ma{B}}_k = [\bar{\ma{b}}_{1,k},\ldots,\bar{\ma{b}}_{T,k}]\in\bb{C}^{M_R\times T}$ is the additive noise at the receiver modeled as a complex Gaussian random variable with zero mean and unitary variance, i.e., $\sim \mathcal{CN}(\ma{0}, M_R \ma{I}_{M_R})$.

Assuming that the transmitter sends orthogonal pilot sequences, which requires $T\geq M_T$, and after matched-filtering using the known pilots, we get\footnote{{\color{black}According to \cite{Clerckx_CL_EA_2024}, when taking into account the mutual coupling between BD-RIS elements, the term $\ma{S}_k$ in (13) is replaced by $(\ma{S}_k^{-1} - \ma{S}_{II})^{-1}$, where $\ma{S}_{II}$ denotes the mutual coupling matrix. This means the whole channel estimation scheme will still work by regarding the term $(\ma{S}_k^{-1} - \ma{S}_{II})^{-1}$ as a whole, while the design of $\ma{S}_k$ should be carefully re-considered to capture the impact induced by $\ma{S}_{II}$}. In this work, we have neglected the coupling effects in BD-RIS since the study of this aspect is out of the scope of this paper. {\color{black} This assumption can be achieved when 1) the inter-element spacing is sufficiently large (at least half-wavelength), and 2) the antenna elements in BD-RIS are perfectly matched \cite{Nerini24TWC}. This allows us to have $\ma{S}_{II} = \ma{0}$, leading to a linear channel model widely adopted in the RIS literature.}}
\begin{equation}
    \label{eq:rec_sig_k}\ma{Y}_k= \ma{G}\ma{S}_k\ma{H}^{\text{T}}+ \ma{B}_k \in \bb{C}^{M_R \times T},
\end{equation}
where $\ma{Y}_k= \bar{\ma{Y}}_k\ma{X}^{\text{H}}$ and $\ma{B}_k=\bar{\ma{B}}_k\ma{X}^{\text{H}}$ are the filtered pilot signals and noise. 
Considering a group-connected \ac{BD-RIS} architecture \cite{shen2021modeling}, the $N$ reflecting elements are divided into $Q$ groups, each with $\bar{N}$ elements connected, i.e., $N= \bar{N} \cdot Q$. In this case, the \textcolor{black}{scattering matrix} is expressed as $\ma{S}_k = \text{blkdiag}(\ma{S}^{(1)}_{k}, \ldots \ma{S}^{(Q)}_{k}) \in \bb{C}^{N \times N}$, where 
the $q$-th matrix $\ma{S}^{(q)}_{k} \in \bb{C}^{\bar{N} \times \bar{N}}$ satisfies $\ma{S}^{(q)\text{H}}_{k}\ma{S}^{(q)}_{k} = \ma{I}_{\bar{N}}$. Hence, (\ref{eq:rec_sig_k}) translates into a sum of $Q$ blocks  
\begin{equation}
    \label{eq:rec_sig_k_sum} \ma{Y}_k = \sumx{q=1}{Q} \ma{G}^{(q)}\ma{S}^{(q)}_{k}\ma{H}^{(q)\text{T}} + \ma{B}_k \in \bb{C}^{M_R \times T},
\end{equation}
\textcolor{black}{
where $\ma{H}^{(q)} \in \bb{C}^{M_T \times \bar{N}}$ and $\ma{G}^{(q)}\in \bb{C}^{M_R \times \bar{N}}$ correspond to the $q$-th block of  $\ma{H}\in \bb{C}^{M_T \times \bar{N}Q}$ and  $\ma{G} \in \bb{C}^{M_R \times \bar{N}Q}$, respectively, defined as follows
\begin{align}
    \ma{H}^{(q)} &= \ma{H}_{.,\, [(q-1)\bar{N}+1,\ldots, q\bar{N}]} \in \bb{C}^{M_T \times \bar{N}}, \, q = 1,\ldots,Q,\\ 
    \ma{G}^{(q)} &= \ma{G}_{.,\, [(q-1)\bar{N}+1,\ldots, q\bar{N}]} \in \bb{C}^{M_R \times \bar{N}},\, q =1,\ldots,Q.
\end{align}
Hence, the channel matrices can be seen as a concatenation of smaller submatrices such that $\ma{H}=[\ma{H}^{(1)},\ldots, \ma{H}^{(Q)}] \in \bb{C}^{M_T \times \bar{N}Q}$ and $\ma{G}=[\ma{G}^{(1)},\ldots, \ma{G}^{(Q)}] \in \bb{C}^{M_R \times \bar{N}Q}$.}

\subsection{Least squares channel estimation}\label{secIIIb}
We start by recalling the conventional LS channel estimation as a reference for the proposed solutions. Defining $\ma{y}_{k} = \vec{\ma{Y}_k} \in \bb{C}^{M_RM_T \times 1}$, and using (\ref{prop:vec}), the noiseless vectorized received signal at the $k$-th block can be expressed as
\begin{align}
  \notag \ma{y}_{k}   &=  \textrm{vec}\Big(\sumx{q=1}{Q} \ma{G}^{(q)}\ma{S}^{(q)}_k\ma{H}^{(q)\text{T}}\Big)\hspace{-0.05cm}= \hspace{-0.05cm}\sumx{q=1}{Q} \big(\ma{H}^{(q)} \hspace{-0.05cm}\otimes\hspace{-0.05cm}\ma{G}^{(q)}\big) \vec{\ma{S}^{(q)}_{k}} \\
 \notag  
\label{eq:approach_1} &=  (\ma{H} \blockron \ma{G})\vec{\ma{\bar{S}}_k}, 
\end{align}
where $\ma{\bar{S}}_k = [\vec{\ma{S}^{(1)}_{k}},\ldots,\vec{\ma{S}^{(Q)}_{k}}] \in \bb{C}^{\bar{N}^2 \times Q}$, and
\begin{equation}
\ma{H} \blockron \ma{G}=\big[\ma{H}^{(1)}  \hspace{-0.05cm}\otimes\hspace{-0.05cm}\ma{G}^{(1)},\ldots,\ma{H}^{(Q)} \hspace{-0.05cm} \otimes\hspace{-0.05cm}\ma{G}^{(Q)}\big] \in \bb{C}^{M_RM_T \times \bar{N}^2Q}, \nonumber
\end{equation}
is the composite block-Kronecker-structured MIMO channel matrix that concatenates the composite channels associated with the $Q$ BD-RIS groups. Defining $\ma{T}\doteq \ma{H} \blockron \ma{G}$ and $\ma{\bar{S}} = [\vec{\ma{\bar{S}}_{1}},\ldots,\vec{\ma{\bar{S}}_{K}}]^{\text{T}}\in \bb{C}^{ K \times \bar{N}^2Q}$ and collecting the received signal over $K$ blocks we have
\begin{align}\label{eq:sig_model_noisy}
\ma{Y} &= [\ma{y}_1, \ldots, \ma{y}_K]=  \ma{T}\ma{\bar{S}}^{\text{T}} + \ma{B}\in \bb{C}^{M_RM_T \times K}, 
\end{align}
where $\ma{B}$ is the corresponding noise term.
An estimate of the composite channel $\ma{T}$ can be obtained by right-filtering using the known BD-RIS training matrix, i.e., $\hat{\ma{T}} = \ma{Y}(\ma{\bar{S}}^{\text{T}})^{\dagger}$
as a solution to the following \ac{LS} problem
\begin{align}
  \label{eq:ls_problem}  \hat{\ma{T}} = \underset{\ma{T}}{\argmin} \big\| \ma{Y} - \ma{T}\ma{\bar{S}}^{\text{T}}\big\|^2_F,
\end{align} 
where $\hat{\ma{T}} \approx \hat{\ma{H}} \blockron \hat{\ma{G}}$ is an estimate of the composite channel. Note that assuming an orthogonal design for \ac{BD-RIS} training matrix $\ma{\bar{S}} \in \mathbb{C}^{K \times \bar{N}^2Q}$, the solution is found by simplified matched filtering \cite{Clerckx_Arxiv_2024}.
In this case, the estimate of the composite channel can also be found as $\hat{\ma{T}}= \ma{Y}\ma{\bar{S}}^{\ast}$.
The LS solution requires $K \geq \bar{N}^2Q$ to ensure a unique estimation of the composite channel. This constraint may be too restrictive, especially for a moderate number of scattering elements, due to the quadratic dependency on the number of connected BD-RIS elements in each group.

\color{black}{
\subsection{Motivation}\label{sec_baseline_design}

\textcolor{black}{
Although the design proposed in \cite{Clerckx_Arxiv_2024} is optimal in the mean square error (MSE) sense, \ac{LS} channel estimation based on (\ref{eq:ls_problem}) ignores the built-in block Kronecker structure of the effective MIMO channel since only an estimate of the ``composite'' channel is obtained. Hence, it requires a significant training overhead with a quadratic growth with the group size $\bar{N}$. As discussed in the following sections, the block Kronecker product structure linking the involved channel matrices can be efficiently exploited, allowing us to obtain enhanced channel estimates compared to the baseline LS solution. Additionally, it turns out that the received signal in (\ref{eq:rec_sig_k_sum}) can be recast using a tensor modeling approach. This is possible by reformulating the training design and rebuilding the BD-RIS training structure as a third-order tensor. Then, by resorting to tensor decomposition algorithms and capitalizing on their intrinsic uniqueness properties, decoupled estimates of the individual channels $\ma{G}$ and $\ma{H}$ can be obtained with a significantly reduced training overhead and improved accuracy. Additionally, as will be shown later (Section \ref{Sec:bdris_tensor_design}), the proposed BD-RIS training design is flexible to allow operation under more challenging system setups with $K << \bar{N}^2Q$, which have not yet been considered in the literature. 
}

{\color{black}
\section{Tensor signal modeling} \label{Sec:Tensor_formulation}
In this section, we recast the signal model for the received pilots using a tensor decomposition approach. In correspondence with the background material presented in Section \ref{Sec:Tensor_background}, we start by formulating the tensor signal model for fully connected architectures. Then, we discuss the tensor signal formulation for the group-connected case, which is of practical interest due to its lower implementation complexity. The main expressions derived in this section will be exploited later to derive the proposed channel estimation methods.

\subsection{Fully connected case}
Starting from the signal model in equation (\ref{eq:rec_sig_k}) and omitting the noise term for notation convenience, the received pilot signal at the $k$-th block can be expressed as
\begin{equation}
    \label{eq:rec_sig_k_sum2} \ma{Y}_k =\ma{G}\ma{S}_{k}\ma{H}^{\text{T}} \in \bb{C}^{M_R \times M_T}.
\end{equation}

By analogy with equation (\ref{eq:tucker_2_slice}), we can
interpret the $k$-th received pilot signal matrix as the $k$-th frontal slice of the \textit{received pilot tensor} $\ten{Y} \in \bb{C}^{M_R \times M_T \times K}$ that follows a Tucker-2 decomposition \cite{Kolda2009,FavierAlmeida2014}. Using (\ref{eq:tenZ_tucker_2}) with the correspondences $(\ma{A},\ma{B},\ma{C},\ten{G}) \leftrightarrow (\ma{G},\ma{H},\ma{I}_K,\ten{S})$, we can express the received pilot signal tensor using the $n$-mode product notation as 
\begin{align}\label{eq:t2_bdris}
\ten{Y}  = \ten{S}  \times_1 \ma{G} \times_2 \ma{H}, 
\end{align}
where $\ten{Y} = \ma{Y}_{1} \sqcup_3 \ma{Y}_{2} \sqcup_3 \ldots \sqcup_3 \ma{Y}_{K}  \in \bb{C}^{M_R\times M_T \times K}$, and $\ten{S} = \ma{S}_{1} \sqcup_3 \ma{S}_{2} \sqcup_3 \ldots \sqcup_3 \ma{S}_{K} \in \bb{C}^{\bar{N} \times \bar{N} \times K}$. Note that $\ten{S}$ results from concatenating the BD-RIS scattering matrices along the third dimension. We refer to $\ten{S}$ as the \textit{BD-RIS training tensor}.

For the fully-connected case and adopting the Tucker representation in (\ref{eq:t2_bdris}), in correspondence with (\ref{eq:tucker_2_u1}), (\ref{eq:tucker_2_u2}), and (\ref{eq:tucker_2_u3}), we can deduce the following matrix unfoldings for the received pilot signal tensor:
\begin{align}
    \label{eq:tenY_1_mod} 
        \nmode{Y}{1} &=   \ma{G}[\ten{S}]_{(1)}\left( \ma{I}_K \otimes \ma{H} \right)^{\text{T}} \in \bb{C}^{M_R \times M_TK},  \\
    \label{eq:tenY_2_mod} 
     \nmode{Y}{2} &=   \ma{H}[\ten{S}]_{(2)}\left(\ma{I}_K
 \otimes \ma{G} \right)^{\text{T}} \in \bb{C}^{M_T \times M_RK},    \\
    \label{eq:tenY_3_mod} \nmode{Y}{3} &= [\ten{S}]_{(3)} (\ma{H} \otimes \ma{G})^{\text{T}}\in \bb{C}^{K \times M_RM_T},
\end{align}
where $[\ten{S}]_{(n)}$ is the $n$-mode unfolding of the BD-RIS training tensor, $n=1,2,3$.
Following (\ref{eq:nmode_1}), (\ref{eq:nmode_2}), and (\ref{eq:nmode_3}), these matrix unfoldings are respectively given by
\begin{eqnarray}
&&\hspace{-2ex}\nmode{S}{1} = [\ten{S}_{..1},\ldots,\ten{S}_{..K}] \in \bb{C}^{N \times NK},\label{slicesk_1}\\
&&\hspace{-2ex} \nmode{S}{2} = [\ten{S}_{..1}^{\text{T}},\ldots,\ten{S}_{..K}^{\text{T}}] \in \bb{C}^{N \times NK},\label{slicesk_2}\\
&&\hspace{-2ex}\nmode{S}{3} = [\text{vec}(\ten{S}_{..1}),\ldots,\text{vec}(\ten{S}_{..K})]^{\text{T}}  \in \bb{C}^{K \times N^2},\label{slicesk_3}
\end{eqnarray}
where $\ten{S}_{..k} \in \bb{C}^{N \times N}$, the $k$-th frontal slice of the tensor $\ten{S} \in \mathbb{C}^{N \times N \times K}$, corresponds to the BD-RIS scattering matrix associated with the $k$-th training block, $k=1,\ldots, K$.

\subsection{Group-connected case}
Let us now consider the group-connected architecture. Starting from the signal model in (\ref{eq:rec_sig_k_sum}), and omitting the noise term, we have
\begin{equation}
    \label{eq:rec_sig_k_sum2} \ma{Y}_k = \sumx{q=1}{Q} \ma{G}^{(q)}\ma{S}^{(q)}_{k}\ma{H}^{(q)\text{T}}.
\end{equation}
In this case, the received pilot tensor $\ten{Y} \in \bb{C}^{M_R \times M_T \times K}$ corresponds to a block Tucker-2 decomposition, i.e., we can rewrite (\ref{eq:t2_bdris}) as a sum of $Q$ tensor blocks
\begin{align}\label{eq:tenY}  
\ten{Y}  = \sumx{q=1}{Q}  \ten{S}^{(q)}  \times_1 \ma{G}^{(q)} \times_2 \ma{H}^{(q)}, 
\end{align}
where $\ten{S}^{(q)} \in \mathbb{C}^{\bar{N} \times \bar{N} \times K}$ is the BD-RIS training tensor associated with the $q$-th group.

Group-connected architectures are commonly adopted due to implementation complexity. In this case, the BD-RIS training tensor is ``sparse'' due to its block-diagonal structure. Figure \ref{fig:BT_sig} illustrates the decomposition of received pilot signals in tensor form, corresponding to a Tucker-2 decomposition composed of $Q$ blocks. This decomposition can also be viewed as a special block-term decomposition (BTD), more specifically, also referred to as a ``type-2 BTD'', which represents a tensor into a sum of rank-$(\bar{N},\bar{N},\cdot)$ tensor blocks \cite{Lathauwer18blockII,Lathauwer18blockIII}. From this figure, we can see that the assumption of a group-connected architecture implies a BD-RIS training tensor having a ``block-diagonal'' structure, the $k$-th frontal slice of which corresponds to the BD-RIS training matrix associated with the $k$-th block. Note that a fully connected BD-RIS architecture is equivalent to having a fully dense BD-RIS training tensor in Figure \ref{fig:BT_sig}, which does not have zero off-diagonal blocks. Although we focus on the group-connected case based on (\ref{eq:tenY}), any (sub-connected) architecture can be captured by the general tensor representation in (\ref{eq:t2_bdris}), the difference being in the structure of the BD-RIS training tensor $\ten{S}$. 

Similarly to (\ref{slicesk_1})-(\ref{slicesk_3}), we construct the unfoldings of the $q$-th BD-RIS scattering tensor as
\begin{eqnarray}
&&\hspace{-7ex}[\ten{S}^{(q)}]_{(1)}=[\ten{S}^{(q)}_{..1},\ldots,\ten{S}^{(q)}_{..K}] \in \bb{C}^{\bar{N} \times \bar{N}K},\label{eq:Sq_unf1}\\
&&\hspace{-7ex}[\ten{S}^{(q)}]_{(2)}=[\ten{S}^{^{(q)}\text{T}}_{..1},\ldots,\ten{S}^{^{(q)}\text{T}}_{..K}] \in \bb{C}^{\bar{N} \times \bar{N}K},\label{eq:Sq_unf2}\\
&&\hspace{-7ex}[\ten{S}^{(q)}]_{(3)}= [\text{vec}(\ten{S}^{(q)}_{..1}),\ldots,\text{vec}(\ten{S}^{(q)}_{..K})]^{\text{T}}  \in \bb{C}^{K \times \bar{N}^2}.\label{eq:Sq_unf3}
\end{eqnarray}
{\color{black} Hence, equivalent expressions can be obtained for the group-connected case for the 1-mode and 2-mode matrix unfoldings of the received pilot tensor, which can be obtained by rewriting (\ref{eq:tenY_1_mod}) and (\ref{eq:tenY_2_mod}) as a sum of $Q$ blocks
\begin{eqnarray}
&&\hspace{-8ex}\nmode{Y}{1}= [\ma{G}^{(1)}, \ldots, \ma{G}^{(Q)}]\cdot \left[\begin{array}{c} \hspace{-1ex} [\ten{S}^{(1)}]_{(1)}(\ma{I}_K \otimes \ma{H}^{(1)\text{T}}) \hspace{-1ex} \\ \hspace{-1ex} \vdots \hspace{-1ex} \\ \hspace{-1ex} [\ten{S}^{(Q)}]_{(1)}(\ma{I}_K \otimes \ma{H}^{(Q)\text{T}})\hspace{-1ex} 
\end{array}\right]\label{eq:unf1block}\\
&&\hspace{-8ex}\nmode{Y}{2}= [\ma{H}^{(1)}, \ldots, \ma{H}^{(Q)}]\cdot \left[\begin{array}{c} \hspace{-1ex} [\ten{S}^{(1)}]_{(2)}(\ma{I}_K \otimes \ma{G}^{(1)\text{T}}) \hspace{-1ex} \\ \hspace{-1ex} \vdots \hspace{-1ex} \\ \hspace{-1ex} [\ten{S}^{(Q)}]_{(2)}(\ma{I}_K \otimes \ma{G}^{(Q)\text{T}})\hspace{-1ex} 
\end{array}\right]\label{eq:unf2block}
\end{eqnarray}
}

Likewise, the 3-mode matrix unfolding of the received pilot tensor can be obtained by rewriting (\ref{eq:tenY_3_mod}) as a sum of $Q$ blocks
\begin{eqnarray}\label{eq:unf_mod_3}
\hspace{-4ex}\nmode{Y}{3}&=&\Big[ [\ten{S}^{(1)}]_{(3)}, \ldots,[\ten{S}^{(Q)}]_{(3)}\Big]\left[\begin{array}{c} \hspace{-1ex} \ma{H}^{(1)\text{T}} \otimes \ma{G}^{(1)\text{T}} \hspace{-1ex} \\ \hspace{-1ex} \vdots \hspace{-1ex} \\ \hspace{-1ex} \ma{H}^{(Q)\text{T}} \otimes \ma{G}^{(Q)\text{T}}\hspace{-1ex} \end{array}\right]\nonumber\\
\hspace{-4ex}&=& \ma{S}_{3}(\ma{H} \blockron \ma{G})^{\text{T}},
\end{eqnarray}
where 
\begin{equation}\label{eq:compactS3}
\ma{S}_{3}\doteq\big[ [\ten{S}^{(1)}]_{(3)}, \ldots,[\ten{S}^{(Q)}]_{(3)}\big] \in \mathbb{C}^{K\times \bar{N}^2Q},
\end{equation}
is referred to as the \emph{compact 3-mode unfolding} that concatenates the 3-mode unfoldings of the $Q$ tensors along its columns. This matrix is key to deriving our first channel estimation method, which relies on channel separation after the LS estimation step using a Kronecker-product factorization method. Likewise, the 1-mode and 2-mode unfoldings of the received pilot tensor, given by the two expressions in (\ref{eq:unf1block})-(\ref{eq:unf1block}), are the basis for the formulation of the second algorithm, which yields direct decoupled estimations of the two involved channels using an iterative algorithm, as will be shown later. 

\begin{figure}[!t]
	\centering\includegraphics[scale=0.31]{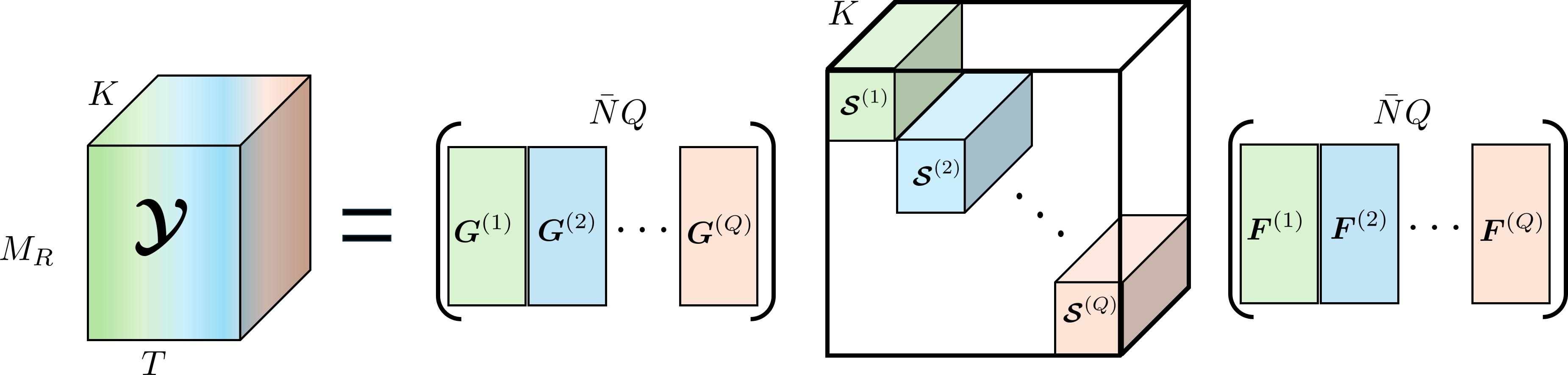}
	\caption{\small \textcolor{black}{Illustration of the decomposition of the noiseless 3-D received signal tensor $\ten Y$ of dimensions $M_R \times T \times K$ for a group-connected BD-RIS communication system. The resulting receiver pilot tensor can be viewed as a sum of $Q$ tensor blocks. The full BD-RIS training tensor is given by a block-diagonal concatenation of $Q$ component tensors $\ten S^{(q)}$, $q=1, \ldots, Q$, each of dimensions $\bar{N} \times \bar{N} \times K$.}}
	\label{fig:BT_sig}
\end{figure}

\textit{Remark 1}: One should note that the compact 3-mode unfolding expression in (\ref{eq:compactS3}) is preferable to the one in (\ref{eq:tenY_3_mod}) when considering a group-connected architecture. This is because the 3-mode unfolding of the BD-RIS tensor $\ten S \in \mathbb{C}^{\bar{N}Q \times \bar{N}Q  \times K}$ is sparse due to its block-diagonal structure (see Figure \ref{fig:BT_sig}). On the other hand, the compact 3-mode unfolding $\ma{S}_{3}$ concatenates the 3-mode unfoldings of $Q$ tensor blocks. 
While $[\ten S]_{(3)}$ is a sparse $K \times \bar{N}^2Q^2$ matrix with a $Q$ times factor increase on the number of columns, $\ma{S}_3$ is a dense $K \times \bar{N}^2Q$ matrix. {\color{black} Adopting the latter yields a more compact estimation of the composite channel in the group-connected case as the block Kronecker product of each group's component.}
Hence, we adopt the compact 3-mode expression in (\ref{eq:compactS3}) to formulate the block Tucker Kronecker factorization (BTKF) algorithm for channel estimation.

\textit{Remark 2}: The tensor model for single-connected RIS corresponds to a special case of (\ref{eq:t2_bdris}), where the $K$ frontal slices $\ten S_{..1}, \ldots, \ten{S}_{..K}$ of the scattering tensor $\ten{S} \in \mathbb{C}^{N\times N \times K}$ are diagonal matrices. It can also be seen as a special case of (\ref{eq:tenY}), where $\bar{N}=1$ and $N=Q$. In tensor notation, this is equivalent to the reduced representation $\ten{S} = \ten{I}_{3,N} \times_3 \ma{S}$, where $\ma{S} \in \mathbb{C}^{K \times N}$ is the diagonal RIS training matrix holding the set of $N$ phase shifts of each training block along its $K$ rows. In this case, the tensor signal model (\ref{eq:t2_bdris}) reduces to
$$\ten{Y}= (\ten{I}_{3,N} \times_3 \ma{S}) \times_1 \ma{G} \times_2 \ma{H}= \ten{I}_{3,N} \times_1 \ma{G} \times_2 \ma{H} \times_3 \ma{S},$$
which corresponds to a PARAFAC tensor model for the received pilots with single-connected RIS \cite{gil2021}. 
 {\color{black} Table~\ref{tab:ris_models} summarizes signal models for different architectures: single-connected, group-connected, and fully connected, highlighting their differences, the resulting structure of the composite channel, and the related tensor model.}

\begin{table*}[]
\renewcommand{\arraystretch}{1.75}
\centering
\caption{{\color{black} Comparison of signal models for different architectures: single-connected, group-connected, and fully connected.}}
\label{tab:ris_models}
\resizebox{\textwidth}{!}{ 
\begin{tabular}{|l|l|l|l|}
\hline
Received signal (noise-free)&
  \begin{tabular}[c]{@{}l@{}}Composite channel 
  \end{tabular}  &
  Connection degree ($N= \bar{N}Q$)&
Tensor model \\ \hline
 $\ma{Y}_{k} = \ma{G}\text{diag}_k(\ma{S})\ma{H}^{\text{T}}$ & $\ma{H} \diamond \ma{G} \in \bb{C}^{M_RM_T \times N}$   & Single-connected $\bar{N}=1$, $Q = N$  &  PARAFAC \cite{gil2021}\\ \hline
  $\ma{Y}_{k} = \sumx{q=1}{Q}\ma{G}^{(q)}\ma{S}_k^{(q)}\ma{H}^{(q)\text{T}}$& $\ma{H} \blockron \ma{G} \in \bb{C}^{M_RM_T \times \bar{N}^2Q}$   & Group-connected, $\bar{N}\geq 2, Q \geq 2$ & Block Tucker-$2$ \\  \hline
 $\ma{Y}_{k} = \ma{G}\ma{S}_k\ma{H}^{\text{T}}$& $\ma{H} \otimes \ma{G} \in \bb{C}^{M_RM_T \times N^2}$ & Fully-connected $\bar{N}=N,Q=1$  &  Tucker-$2$ \\ \hline
\end{tabular}
}
\end{table*}

\section{Tensor-Based channel estimation methods} \label{Sec:TB_CE_methods}
In this section, we formulate the proposed channel estimation methods by capitalizing on the tensor signal structures discussed in the previous section. Two algorithms are proposed to solve the channel estimation problem \emph{via} decoupling the estimates of the involved channel matrices $\ma{G}$ and $\ma{H}$. The first one, BTKF, is a closed-form solution that extracts the channel estimates via a block Kronecker factorization problem, which boils down to solving a set of rank-one matrix approximation problems. The second is BTALS, which directly estimates the channel matrices using an iterative procedure. We also discuss identifiability and uniqueness issues and their overall implications for system design. The design of the BD-RIS training tensor is a separate topic that will be addressed in the next section.

\subsection{Block Tucker Kronecker factorization algorithm (BTKF) algorithm for decoupled channel estimation}  \label{Sec:BTKF}
Taking into consideration the noise term, we recall the received signal model in tensor form as
\begin{align}\label{eq:t2_bdris_noise}
\ten{Y}  = \ten{S}  \times_1 \ma{G} \times_2 \ma{H} \,+ \ten{B}, 
\end{align}
where $\ten{B} \in \mathbb{C}^{M_R \times M_T \times K}$ is the additive noise tensor.

Closed-form estimates of $\ma{H}$ and $\ma{G}$ can be obtained by exploiting the block Kronecker structure of the compact $3$-mode unfolding of the received pilot signal tensor in (\ref{eq:unf_mod_3}), which is given by $[\ten{Y}]_{(3)}=\ma{S}_{3}(\ma{H} \blockron \ma{G})^{\text{T}} + [\ten B]_{(3)}$. First, assuming $\ma{S}_{3}$ has full column rank, right-filtering gives
\begin{align}
  \label{eq:block_kronencker} \ma{Z} =  
\big(\ma S_{3}^{\dagger}\nmode{{Y}}{3}\big)^{\text{T}} 
  &\approx \ma{H}  \blockron \ma{G}  \in \bb{C}^{M_RM_T \times \bar{N}^2Q}. 
\end{align}

To obtain decoupled estimates of the involved channel matrices $\ma{H}$ and $\ma{G}$ from the filtered signal in (\ref{eq:block_kronencker}), we formulate the following optimization problem
\begin{align}
\label{prob:ls_kron}    \{\hat{\ma{H}}, \hat{\ma{G}}\} = \underset{\ma{H},\ma{G}}{\argmin} \fronorm{\ma{Z} - \ma{H} \blockron \ma{G}}^2.
    \end{align}
By making use of the partition of the BD-RIS phase shifts into $Q$ groups, one can easily note that
\begin{equation}\label{eq:blockZ}
[\ma{Z}^{(1)},\ldots,\ma{Z}^{(Q)}]  \approx \big[\ma{H}^{(1)}  \hspace{-0.05cm}\otimes\hspace{-0.05cm}\ma{G}^{(1)},\ldots,\ma{H}^{(Q)} \hspace{-0.05cm} \otimes\hspace{-0.05cm}\ma{G}^{(Q)} \big],   
\end{equation}
where $\ma{Z}^{(q)} = \ma{Z}_{., [(q-1)\bar{N}^2 + 1,\ldots, q\bar{N}^2]}$ is the $q$-th block matrix of the filtered signal, $q=1,\ldots, Q$.
From such a block structure, we can recast this problem as $Q$ independent sub-problems executed in parallel, with the $q$-th problem being defined as
  \begin{align}
\label{prob:ls_kron_Q} \{\hat{\ma{H}}^{(q)}, \hat{\ma{G}}^{(q)}\} = \underset{\ma{H}^{(q)},\ma{G}^{(q)}}{\argmin} \fronorm{\ma{Z}^{(q)}- \ma{H}^{(q)} \otimes \ma{G}^{(q)}}^2, 
\end{align} 
$q=1, \ldots, Q$. \textcolor{black}{This problem can be solved using the classical nearest Kronecker approximation method originally proposed in \cite{vanloan}}, a closed-form approach that yields the best factorization of the Kronecker product of two matrices.

Specifically, by properly permuting the elements of $\ma{Z}^{(q)}$, problem (\ref{prob:ls_kron_Q}) can be recast as a simple rank-one matrix approximation, such that it can be rewritten as
\begin{align}
   \label{prob:ls_kron_Q_rank_one}   \{\hat{\ma{h}}^{(q)}, \hat{\ma{g}}^{(q)}\} = \underset{\ma{h}^{(q)},\ma{g}^{(q)}}{\argmin} \fronorm{\overline{\ma{Z}}^{(q)} - \ma{g}^{(q)} \ma{h}^{(q)\text{T}}}^2,
\end{align}
where $\overline{\ma{Z}}^{(q)} \in \bb{C}^{M_R \bar{N} \times M_T\bar{N}}$ is a reshaped version of $\ma{Z}^{(q)} \in \bb{C}^{M_RM_T  \times \bar{N}^2}$, $\ma{g}^{(q)} = \vec{\ma{G}^{(q)}} \in \bb{C}^{M_R\bar{N} \times 1}$, and $\ma{h}^{(q)} = \vec{\ma{H}^{(q)}} \in \bb{C}^{M_T\bar{N} \times 1}$. Please note that a rank-one matrix approximation is unique to a scaling factor affecting the estimated vectors. Thus, defining $\overline{\ma{Z}}^{(q)} = \ma{U}^{(q)}\ma{\Sigma}^{(q)}\ma{V}^{(q)\text{H}}$ the estimates of $\ma{g}^{(q)}$ and $\ma{h}^{(q)}$ correspond respectively to the dominant left and right singular vectors $\ma{u}^{(q)}_{1} = \ma{U}^{(q)}_{.1}\in \bb{C}^{M_R\bar{N} \times 1}$ and $\ma{v}^{*(q)}_{1} = \ma{V}^{*(q)}_{.1}\in \bb{C}^{M_T\bar{N} \times 1}$ associated with the first singular value $\sigma_{1}^{(q)} = \ma{\Sigma}^{(q)}_{1,1}$, i.e.,

\begin{align}
\label{eq:ls_kron_sol_rank_1_G} \hat{\ma{g}}^{(q)} &= \sqrt{\sigma_1^{(q)}}\ma{u}^{(q)}_1, \quad \hat{\ma{G}}^{(q)} = \unvec{\hat{\ma{g}}^{(q)}}_{M_R \times \bar{N}}, \\
\label{eq:ls_kron_sol_rank_1_H} \hat{\ma{h}}^{(q)} &= \sqrt{\sigma_1^{(q)}}\ma{v}^{*(q)}_1, \quad \hspace{-0.08cm}\hat{\ma{H}}^{(q)} = \unvec{\hat{\ma{h}}^{(q)}}_{M_T \times \bar{N}}." 
\end{align}}
The global estimates of the channel matrices are then formed by collecting the $Q$ estimated blocks
\begin{align}
    \hat{\ma{G}} &= [\hat{\ma{G}}^{(1)},\ldots,\hat{\ma{G}}^{(Q)}]  \in \bb{C}^{M_R \times \bar{N}Q},\\
    \hat{\ma{H}} &= [\hat{\ma{H}}^{(1)},\ldots,\hat{\ma{H}}^{(Q)}]   \in \bb{C}^{M_T \times \bar{N}Q}.
\end{align}

\subsection{Block Tucker alternating least squares (BTALS) algorithm for decoupled channel estimation} \label{Sec:BTALS}

Starting from the noisy received pilot model in the tensor form shown in equation (\ref{eq:t2_bdris_noise}), let us consider the following problem
\begin{equation}
\underset{\mathbf{G},\mathbf{H}}{\text{min}}\Big\|\ten{Y} - \ten{S} \times_1 \ma{G} \times_2 \ma{H} \Big\|_{\text{F}}^{2}.
\label{optpro}
\end{equation}
This problem is nonlinear in the unknowns since the solution involves products of the coefficients of the channel matrices $\ma{H}$ and $\ma{G}$. However, from tensor decomposition theory \cite{comon2009tensor}, this problem can be efficiently solved through an \acf{ALS} estimation algorithm \cite{comon2009tensor,Bro1}, a popular iterative method for fitting a tensor model thanks to its simplicity and monotonic convergence property. In our case, the algorithm yields decoupled estimates of the channel matrices by converting the problem in (\ref{optpro}) into two linear LS sub-problems that are solved iteratively. 

The following developments consider the group-connected case due to its practical interest.\footnote{The final expressions for the fully connected case can be easily deduced from the group-connected one by simply assuming $Q=1$ and $\bar{N}=N$.} {\color{black} For later convenience, let us rewrite the expressions in (\ref{eq:unf1block})-(\ref{eq:unf1block}) in compact form as 
\begin{eqnarray}
&&\hspace{-4ex}\nmode{Y}{1}= \ma{G}\ma{S}_{1}\ma{T}_{\ma{H}}
\label{eq:unf_mod_12},\quad
\nmode{Y}{2}= \ma{H}\ma{S}_{2}\ma{T}_{\ma{G}}, 
\end{eqnarray}
where $\ma{S}_n\doteq \text{blkdiag}\big([\ten{S}^{(1)}]_{(n)}, \ldots,[\ten{S}^{(Q)}]_{(n)}\big) \in \mathbb{C}^{\bar{NQ} \times \bar{N}KQ}$
collects the $n$-mode unfoldings of the BD-RIS tensor, $n=1,2$, while $\ma{T}_{\ma{H}}$ and $\ma{T}_{\ma{G}}$ are the Kronecker structured matrices
\begin{eqnarray}
&&\hspace{-7ex} \ma{T}_{\ma{H}}\doteq \big[\ma{I}_K \otimes \ma{H}^{(1)},\ldots, \ma{I}_K \otimes \ma{H}^{(Q)} \big]^{\text{T}} \in \mathbb{C}^{\bar{N}KQ\times M_TK }\label{eq:Th},\\
&&\hspace{-7ex} \ma{T}_{\ma{G}}\doteq  \big[\ma{I}_K \otimes \ma{G}^{(1)},\ldots, \ma{I}_K \otimes \ma{G}^{(Q)} \big]^{\text{T}} \in \mathbb{C}^{\bar{N}KQ \times M_RK} \label{eq:Tg}.
\end{eqnarray}

These expressions are exploited to directly estimate $\ma{G}$ and $\ma{H}$
by iteratively solving the following two \ac{LS} problems 
\begin{eqnarray}\label{eq:cost_G}
     \hat{\ma{G}}\hspace{-0.08cm} &= \hspace{-0.08cm}\underset{\ma{G}}{\argmin}\hspace{-0.05cm}\fronorm{\nmode{Y}{1} \hspace{-0.05cm}- \hspace{-0.05cm}\ma{G} \ma{S}_{1}\ma{T}_{\ma{H}}}^2, \\
    \label{eq:cost_H} \hat{\ma{H}}\hspace{-0.08cm} &= \hspace{-0.08cm}\underset{\ma{H}}{\argmin}\hspace{-0.05cm}\fronorm{\nmode{Y} {2} \hspace{-0.05cm}- \hspace{-0.05cm}\ma{H} \ma{S}_{2}\ma{T}_{\ma{G}}}^2, 
\end{eqnarray}
the solutions of which are respectively given by
\begin{align}
     \label{eq:ls_GH_sol}  \hat{\ma{G}} = \nmode{Y}{1}\left[\ma{S}_{1}\ma{T}_{\ma{H}}\right]^{+},\quad 
\hat{\ma{H}} = \nmode{Y}{2}\left[\ma{S}_{2}\ma{T}_{\ma{G}}\right]^+\hspace{-0.15cm}.
\end{align}
The algorithm intertwines the LS estimates of $\ma{G}$ and $\ma{H}$. At each iteration, one matrix is updated based on a previously obtained estimate of the other matrix. This procedure is repeated until convergence. An error measure computed at the end of the $i$-th iteration is given by   
\begin{align}
\label{eq:error_BTALS} \epsilon_{(i)} \doteq \frac{\fronorm{\nmode{Y}{3} -\hat{\ma Y}_{(i)} }^2}{\fronorm{\nmode{Y}{3}}^2},
\end{align}
where $\hat{\ma Y}_{(i)} \doteq [\ten{S}]_{(3)}\big(\ma{\hat{H}}_{(i)} \otimes \ma{\hat{G}}_{(i)}\big)^{\text{T}}$ is the reconstructed signal model at the $i$-th iteration.
The convergence is declared when the error does not significantly change between two successive iterations, which implies that $|\epsilon_{(i)} - \epsilon_{(i-1)}| \leq \eta$, where $\eta=10^{-6}$ is a convergence threshold. A summary of the steps of the proposed BALS algorithm is provided in Algorithm~\ref{algorithm_bals}.}

Such an alternating estimation procedure may improve or maintain but cannot worsen the current fit, usually leading to a global minimum solution \cite{Sidiropoulos_DSCDMA}. For simplicity, random initialization can be adopted. However, one may resort to enhanced initialization and acceleration procedures \cite{Sidiropoulos_DSCDMA,comon2009tensor}. Other initialization strategies that could be used to exploit the structure of the channel matrices are possible, although not advocated here. In our case, however, where the core tensor (represented by the BD-RIS structure) is known at the receiver, such enhancements are not required, and random initialization is enough to obtain satisfactory results.

 \begin{algorithm}[!t]
	\begin{algorithmic}[1]
		\caption{Block Tucker Kronecker factorization algorithm (BTKF) for decoupled channel estimation}\label{algorithm_kron_fac}
		\State \textbf{Inputs}: Received signal tensor $\ten{Y}$ and \ac{BD-RIS} training tensors $\{\ten{S}^{(1)}, \ldots, \ten{S}^{(Q)}\}$.
	\State Compute an estimate of $\ma{Z}$ by filtering
 \begin{equation*}
\big(\ma{S}_{3}^{\dagger}\nmode{{Y}}{3}\big)^{\text{T}}  \approx  \ma{H}  \blockron \ma{G}  \in \bb{C}^{M_RM_T \times \bar{N}^2Q}.
 \end{equation*}
 \For{$q = 1:Q$}	
 \State Partition $\ma{Z}$ into $Q$ matrix blocks 
 $\{\ma{Z}^{(1)}, \ldots, \ma{Z}^{(Q)}\}$:
\begin{equation*}\ma{Z}^{(q)} = \ma{Z}_{.,[(q-1)\bar{N}^2,\ldots,q\bar{N}^2]} \in \bb{C}^{M_RM_T \times \bar{N}^{2}}. \end{equation*}
\State Reshape $\ma{Z}^{(q)}$ to obtain $\overline{\ma{Z}}^{(q)} \in \bb{C}^{M_R\bar{N} \times M_T\bar{N}}$.
\State Define the \ac{SVD} of 
$\overline{\ma{Z}}^{(q)} = \ma{U}^{(q)}\ma{\Sigma}^{(q)}\ma{V}^{(q)\text{H}}$.
\State Obtain an estimate of $\ma{G}^{(q)}$ and $\ma{H}^{(q)}$ as
\begin{align*}
    \hat{\ma{G}}^{(q)} &= \unvec{\sqrt{\sigma_1^{(q)}}\ma{u}^{(q)}_{1}}_{M_R \times \bar{N}}, \\
    \hat{\ma{H}}^{(q)} &= \unvec{\sqrt{\sigma_1^{(q)}}\ma{v}^{*(q)}_{1}}_{M_T \times \bar{N}}.
\end{align*}				
 \EndFor
\State Return $\ma{\hat{G}} = [\ma{\hat{G}}^{(1)},\ldots,\ma{\hat{G}}^{(Q)}]$, $\ma{\hat{H}} =[\ma{\hat{H}}^{(1)},\ldots,\ma{\hat{H}}^{(Q)}]$.
	\end{algorithmic}
\end{algorithm}

 \begin{algorithm}[!t]
	\begin{algorithmic}[1]
		\caption{Block Tucker alternating least squares (BTALS) algorithm for decoupled channel estimation}\label{algorithm_bals}
		\State \textbf{Inputs}: Received signal tensor $\ten{Y}$, \ac{BD-RIS} training tensor $\ten{S}$, 
number $I$ of iterations, and convergence threshold $\eta$.
		\State Set $i=0$. Randomly initialize $\ma{\hat{H}}_{(i=0)}$. 
\For{$i = 1:I$}		
		\State Compute an LS estimate of $\ma{G}_{(i)}$ as
  \begin{eqnarray*}
     &&{\color{black}\ma{T}_{\hat{\ma{H}}(i-1)}\doteq \Big[\ma{I}_K \otimes \ma{H}^{(1)}_{(i-1)},\ldots, \ma{I}_K \otimes \ma{H}^{(Q)}_{(i-1)}\Big]^{\text{T}}} \\
     && \ma{\hat{G}}_{(i)} = \nmode{Y}{1}{\color{black}\left[\ma{S}_{1}\ma{T}_{\hat{\ma{H}}(i-1)}\right]^{+}}.
  \end{eqnarray*}
  	\State Compute an LS estimate of $\ma{H}_{(i)}$ as
   \begin{eqnarray*}
    &&{\color{black}\ma{T}_{\hat{\ma{G}}(i)}\doteq \Big[\ma{I}_K \otimes \ma{G}^{(1)}_{(i)},\ldots, \ma{I}_K \otimes \ma{G}^{(Q)}_{(i)}\Big]^{\text{T}}} \\
   &&\ma{\hat{H}}_{(i)} = \nmode{Y}{2}{\color{black}\left[\ma{S}_{2}\ma{T}_{\hat{\ma{G}}_{(i)}}\right]^+}.
   \hspace{-0.15cm}
   \end{eqnarray*}
   \State 
   Compute $\hat{\ma{Y}}{(i)}= [\ten{S}]_{(3)}\big(\ma{\hat{H}}_{(i)} \otimes \ma{\hat{G}}_{(i)}\big)^{\text{T}}$	
   
   and calculate the error according to (\ref{eq:error_BTALS}).
 
    \State Check convergence and stop if $|\epsilon_{(i)} - \epsilon_{(i-1)}| \leq \eta$.					
 \EndFor
		\State Return $\ma{\hat{G}}_{(i)}$ and $\ma{\hat{H}}_{(i)}$.
	\end{algorithmic}
\end{algorithm}

\textit{Remark 3}: Steps 4 and 5 of Algorithm 2 are based on the compact expressions for the 1-mode and 2-mode unfoldings in (\ref{eq:unf_mod_12}). In the fully connected case, we can alternatively consider the general expressions (\ref{eq:tenY_1_mod})-(\ref{eq:tenY_2_mod}) and replace these updating steps with the $\ma{\hat{G}}_{(i)} = [\ten{Y}]_{(1)}\big[[\ten{S}]_{(1)}(\ma{I}_K \otimes \ma{H}_{(i-1)} )^{\text{T}}\big]^{+}$ and $\ma{\hat{H}}_{(i)} = [\ten{Y}]_{(2)}\big[[\ten{S}]_{(2)}\left(\ma{I}_K \otimes \ma{G}_{(i)} \right)^{\text{T}}\big]^{+}$, respectively, to estimate the channel matrices directly. 

The main steps of the proposed BTKF and BTALS algorithms are summarized in Algorithm \ref{algorithm_kron_fac} and Algorithm \ref{algorithm_bals}.

{\color{black}
\subsection{Uniqueness and scaling ambiguities}\label{subsec:uniqueness}
This section discusses the uniqueness issues related to the estimation of the channel matrices $\ma{G}$ and $\ma{H}$ from the Tucker tensor model using the BTKF and BTALS algorithms. In the following, we obtain applicable conditions involving the training parameters that ensure unique estimates of the channel matrices. First, note that the two algorithms exploit the structure of the received signal tensor differently. While BTKF is based on its 3-mode unfolding, BTALS uses its 1-mode and 2-mode unfoldings. This translates into different requirements involving the system parameters used for channel training.

\emph{Result 1}: Assuming $\textrm{rank}(\ma S_3)=\bar{N}^2Q$, unique solutions for the estimates of the channel matrices $\ma{G}$ and $\ma{H}$ are obtained using the BTKF algorithm according to problem (\ref{prob:ls_kron}). 

\emph{Proof:} Note that left-filtering $[\ten{Y}]_{(3)}$ in (\ref{eq:block_kronencker}) using the 3-mode unfolding of the BD-RIS training tensor $\ma S_3$ yields a unique estimate of the composite channel $\ma{H} \hspace{2pt} \blockron \hspace{2pt}\ma{G}$ if $\ma S_3$ has full column-rank (which requires $K\geq \bar{N}^2Q$). Under this condition, an additional constraint is necessary to extract the individual estimates of the channel matrices, which is accomplished from each block of the estimated composite channel from rank-one matrix approximation steps. $\blacksquare$ 

\emph{Result 2}: Let us assume $\textrm{rank}(\ma S_1)=\textrm{rank}(\ma S_2)=\bar{N}Q$. Uniqueness of the estimates of the channel matrices $\ma{G}$ and $\ma{H}$ according to problems (\ref{eq:cost_G})-(\ref{eq:cost_H}) using the BTALS algorithm requires $KM_T\geq \bar{N}Q$ and $KM_R\geq \bar{N}Q$. 

\emph{Proof:} Let us consider the LS estimation steps in (\ref{eq:ls_GH_sol}). Defining $\ma P_1\doteq \ma{T}_{\ma{H}}\ma{S}_{1}^{\text{T}} \in \mathbb{C}^{KM_T \times \bar{N}Q}$ and $\ma P_2\doteq \ma{T}_{\ma{G}}\ma{S}_{2}^{\text{T}}  \in \mathbb{C}^{KM_R\times \bar{N}Q}$, the uniqueness of $\hat{\ma{G}}=[\ten Y]_{(1)}[\ma P_1^{\text{T}}]^{\dagger}$ and $\hat{\ma{H}}=[\ten Y]_{(2)}[\ma P_2^{\text{T}}]^{\dagger}$ requires that $\textrm{rank}(\ma P_1)=\textrm{rank}(\ma P_2)=N$, so that $\ma P_1$ and $\ma P_2$ left-invertible, which implies $KM_T\geq N$ and $KM_R\geq N$.
Note, however, that these conditions are necessary but not sufficient for uniqueness. Nevertheless, these conditions are useful for the system designer to eliminate invalid system setups that do not yield unique estimates of the channel matrices. Otherwise stated, system setups that violate one of these conditions can be discarded. $\blacksquare$
}

Under the conditions discussed previously, the estimates of $\ma{G}$ and $\ma{H}$ are unique up to scaling ambiguities. To verify this, recall that the received pilot tensor in the noise-free case follows a Tucker-$2$ decomposition (see (\ref{eq:t2_bdris})). Generically, Tucker models are not essentially unique due to rotational freedom since nonsingular transformations compensating each other can be applied to the factor matrix and the core tensor \cite{comon2009tensor,Kolda2009}. Defining non-singular matrices $\ma{T}_{\ma{G}} \in \mathbb{C}^{N \times N}$ and  $\ma{T}_{\ma{H}} \in \mathbb{C}^{N \times N}$, any alternative solution $\hat{\ma{G}}=\ma{G}\ma{T}_{\ma{G}}$, $\hat{\ma{H}}=\ma{H}\ma{T}_{\ma{H}}$ and $\hat{\ten{S}}= \ten{S} \times_1\ma{T}^{-1}_{\ma{G}} \times_2\ma{T}^{-1}_{\ma{H}}$ yields the same result since $\ten{Y}= \hat{\ten{S}} \times_1 \hat{\ma{G}} \times_2 \hat{\ma{H}} = (\ten{S} \times_1\ma{T}^{-1}_{\ma{G}} \times_2\ma{T}^{-1}_{\ma{H}}) \times_1 (\ma{G}\ma{T}_{\ma{G}}) \times_2 (\ma{H}\ma{T}_{\ma{H}}) = \ten{S} \times_1 (\ma{G}(\ma{T}_{\ma{G}}\ma{T}^{-1}_{\ma{G}})) \times_2 (\ma{H}(\ma{T}_{\ma{H}}\ma{T}^{-1}_{\ma{H}}))= \ten{S} \times_1 \ma{G} \times_2 \ma{H}$. In the group-connected case, these nonsingular transformation matrices are confined within each block, and one can also arbitrarily permute the $Q$ blocks without changing the result \cite{Lathauwer18blockII}. However, the core tensor is fixed since the BD-RIS training tensor $\ten{S}$ is known at the receiver. In this case, the only feasible structures for $\ma{T}_{\ma{G}}$ and $\ma{T}_{\ma{H}}$ correspond to scaled identity matrices that compensate each other, i.e., $\ma{T}_{\ma{G}}=\alpha \ma{I}_N$ and $\ma{T}_{\ma{H}}=\beta\ma{I}_N$., with $\alpha\beta=1$. Thus, any alternative solutions for $\hat{\ma{G}}$ and $\hat{\ma{H}}$ satisfy the identities
$$\hat{\ma{G}}=\alpha\ma{G}, \quad \hat{\ma{H}}=\beta\ma{H},\quad \alpha\beta=1.$$
These identities can alternatively be understood by looking at the 3-mode unfolding in (\ref{eq:tenY_3_mod}) and introducing the transformation matrices, yielding
$\nmode{Y}{3} = [\ten{S}]_{(3)} (\hat{\ma{H}} \otimes \hat{\ma{G}})^{\text{T}}=[\ten{S}]_{(3)}((\ma{H}\ma{T}_{\ma{H}}) \otimes (\ma{G}\ma{T}_{\ma{G}}))^{\text{T}}= [\ten{S}]_{(3)}(\ma{T}_{\ma{H}} \otimes \ma{T}_{\ma{G}})^{\text{T}}(\ma{H} \otimes \ma{G})^{\text{T}}$. Note that the only choice for $\ma{T}_{\ma{G}}$ and $\ma{T}_{\ma{H}}$ satisfying the Kronecker-product equation $\ma{T}_{\ma{H}} \otimes \ma{T}_{\ma{G}}=\ma{I}_{N^2}$ are scaled identity matrices that compensate each other.

The reasoning for the group-connected case is similar, but instead applies to each group. Defining the block-diagonal transformation matrices $\ma{T}_G=\textrm{blkdiag}(\ma{T}^{(1)}_G, \ldots, \ma{T}^{(Q)}_G)$ and $\ma{T}_H=\textrm{blkdiag}(\ma{T}^{(1)}_H, \ldots, \ma{T}^{(Q)}_H)$ that act independently over each channel block, and from similar arguments, alternative channel matrices $\hat{\ma{G}}^{(q)}$ and $\hat{\ma{H}}^{(q)}$, associated with the $q$-th BD-RIS group, satisfy the following identities}
\begin{align*}
\hat{\ma{G}}^{(q)}=\alpha^{(q)}\ma{G}^{(q)},\,\,\,\hat{\ma{H}}^{(q)}=\beta^{(q)}\ma{H}^{(q)}, \,\,\,\alpha^{(q)}\beta^{(q)}=1
\end{align*}
for $q = 1,\ldots,Q$.  Hence, individual channel estimation involves a single scalar ambiguity in the fully connected case and $Q$ scalar ambiguities in the group-connected case.

It is worth noting, however, that these scaling ambiguities affecting the individual channels are irrelevant to the optimization of the BD-RIS scattering coefficients and transmit/receive beamforming vectors since they cancel each other when building the composite channel used to optimize the system parameters using state-of-the-art methods \cite{Clerckx_Arxiv_2024}.

\section{BD-RIS training design}\label{Sec:bdris_tensor_design}
{\color{black} As previously discussed, the BTKF and BTALS algorithms exploit the BD-RIS training tensor differently to estimate the individual channel matrices. While BTKF relies on its 3-mode unfolding (\ref{eq:compactS3}) for a two-step channel separation using Kronecker factorization, BTALS exploits its 1-mode and 2-mode unfoldings in (\ref{eq:unf_mod_12}) for an iterative estimation of the channel matrices using alternating least squares.
Proper design of the BD-RIS tensor is crucial to obtaining reasonable estimates of the individual channel matrices using these algorithms.

To address the physical constraints of group-connected BD-RIS, the work \cite{Clerckx_Arxiv_2024} proposed an orthogonal design that requires $K=\bar{N}^2Q$. We propose a BD-RIS training design with a more flexible construction, which covers a broader range of values of $K$, while being valid for smaller training lengths $K<\bar{N}^2Q$, as opposed to the design of \cite{Clerckx_Arxiv_2024}. 
}

{\color{black} Let $\ma{Z}\doteq[\ma{z}_1,\ldots, \ma{z}_{K_1}] \in \mathbb{C}^{ \bar{N}^2\times K_1}$ be as a (possibly truncated) circulant matrix constructed such that $\ma{z}_{k_1} \doteq \textrm{circshift}(\textrm{vec}(\ma{\Omega}_{\textrm{DFT}}),k_1-1)$, $k_1=1, \ldots, K_1$, where $K_1=\textrm{min}(K,\bar{N}^2)$, and $\ma{\Omega}_{\textrm{DFT}}\in \mathbb{C}^{\bar{N} \times \bar{N}}$ is a DFT matrix, with $\ma{\Omega}^{\textrm{H}}_{\textrm{DFT}}\ma{\Omega}_{\textrm{DFT}}=\ma{I}_{\bar{N}}$.
}
Define the tensor $\ten{Z} \in \mathbb{C}^{\bar{N} \times \bar{N} \times K_1}$ whose $k_1$-th frontal slice corresponds to
\begin{equation}\label{eq:Zk1}
\ten Z_{..k_1}\doteq \textrm{unvec}_{\bar{N}\times \bar{N}}(\ma{z}_{k_1}) \in \bb{C}^{\bar{N} \times \bar{N}}, \,\, k_1=1,\ldots, K_1,
\end{equation}
which are column/row-permuted versions of $\ma{\Omega}_{\textrm{DFT}}$.
The basic structure of $\ten{S}^{(q)} \in \bb{C}^{\bar{N} \times \bar{N} \times K}$ has frontal slices given by
\begin{equation}\label{eq:blkdiag_Sk_basis}
\ten{S}^{(q)}_{..k}=[\ma{\Theta}]_{k_2,q}\,\ten{Z}_{..k_1} \, \in \mathbb{C}^{\bar{N}\times \bar{N}},\,\, k\doteq (k_2-1)K_1+k_1
\end{equation}
where $\ma{\Theta}\doteq[\ma{\theta}_1,\ldots, \ma{\theta}_Q]\in \mathbb{C}^{K_2 \times Q}$ is a (possibly truncated) $K_2 \times Q$ Hadamard
or a DFT matrix, where $K_2=K/K_1$ must be an integer. Collecting the $K$ slices in (\ref{eq:blkdiag_Sk_basis}) according to (\ref{eq:Sq_unf1})-(\ref{eq:Sq_unf3}) gives $[\ten{S}^{(q)}]_{(1)}=\ma{\theta}^{\text{T}}_q \otimes [\ten{Z}]_{(1)} \in \mathbb{C}^{\bar{N} \times \bar{N}K}$, $[\ten{S}^{(q)}]_{(2)}=\ma{\theta}^{\text{T}}_q \otimes [\ten{Z}]_{(2)} \in \mathbb{C}^{\bar{N} \times \bar{N}K}$, and $[\ten{S}^{(q)}]_{(3)}=\ma{\theta}_q \otimes [\ten{Z}]_{(3)} \in \mathbb{C}^{K \times \bar{N}^2}$, from which we obtain
\begin{eqnarray}
&&\hspace{-7ex}\ma{S}_{1}\hspace{-.5ex}=\textrm{blkdiag}(\ma{\theta}^{\text{T}}_1 \otimes [\ten{Z}]_{(1)},\ldots, \ma{\theta}^{\text{T}}_Q \otimes [\ten{Z}]_{(1)}),\label{eq:unfoldings_Z1}\\
&&\hspace{-7ex}\ma{S}_{2}\hspace{-.5ex}=\textrm{blkdiag}(\ma{\theta}^{\text{T}}_1 \otimes [\ten{Z}]_{(2)},\ldots, \ma{\theta}^{\text{T}}_Q \otimes [\ten{Z}]_{(2)}),\label{eq:unfoldings_Z2}\\
&&\hspace{-7ex}\ma{S}_{3}\hspace{-.5ex}=\big[\ma{\theta}_1\otimes [\ten{Z}]_{(3)}, \ldots,\ma{\theta}_Q\otimes [\ten{Z}]_{(3)}\big]=\ma{\Theta}\otimes [\ten{Z}]_{(3)},\label{eq:unfoldings_Z3}
\end{eqnarray}
where $[\ten{Z}]_{(1)} \in \mathbb{C}^{\bar{N} \times \bar{N}K_1}$, $[\ten{Z}]_{(2)} \in \mathbb{C}^{\bar{N} \times \bar{N}K_1}$, and $[\ten{Z}]_{(3)} \in \mathbb{C}^{K_1\times \bar{N}^2}$ are the the unfoldings of the circulant tensor $\ten{Z}$ constructed from its frontal slices similarly to (\ref{slicesk_1})-(\ref{slicesk_3}). Since the frontal slices of the tensor $\ten{Z}$ are symmetric due to (\ref{eq:Zk1}), we have $[\ten{Z}]_{(1)}=[\ten{Z}]_{(2)}$, which implies $[\ten{S}^{(q)}]_{(1)}=[\ten{S}^{(q)}]_{(2)}$, $q=1,\ldots, Q$, and, hence, $\ma{S}_{1}=\ma{S}_{2}$ in (\ref{eq:unfoldings_Z1})-(\ref{eq:unfoldings_Z2}). 

\subsection{BD-RIS design for the BTKF algorithm}
The initial step of the BTKF algorithm involves left-filtering the 3-mode unfolding $[\ten Y]_{(3)}$ using the compact 3-mode unfolding $\ma{S}_{3}=\ma{\Theta}\otimes [\ten{Z}]_{(3)}$ of the BD-RIS training tensor.

{\color{black} \emph{Property 1}: If $K_1 =\bar{N}^2$ and $K_2\geq Q$, then $\ma{\Theta}$ and $[\ten{Z}]_{(3)}$ are column-orthogonal matrices
such that $\ma{S}_3$ has full column rank and column-orthogonal, satisfying $\ma{S}^{\text{H}}_3\ma{S}_3=(K/\bar{N})\ma I_{\bar{N}^2Q}$.}

This property can be checked as follows. We first note that $K\geq \bar{N}^2Q$ implies $K_1=\bar{N}^2$ and $K_2=K/\bar{N}^2$. In this case, $[\ten{Z}]_{(3)}$ is a square $\bar{N}^2 \times \bar{N}^2$ circulant matrix with orthogonal rows constructed from circular shifts of the DFT basis in vectorized form, which implies $[\ten{Z}]^{\text{H}}_{(3)}[\ten{Z}]_{(3)}=\bar{N}\ma{I}_{\bar{N}^2}$. From (\ref{eq:unfoldings_Z3}), we have $[\ten{S}^{(q)}]^{\text{H}}_{(3)}[\ten{S}^{(q)}]_{(3)}=(\ma{\Theta}^{\text{H}}\ma{\Theta})\otimes ([\ten{Z}]^{\text{H}}_{(3)}[\ten{Z}]_{(3)})= K/\bar{N}^2 \otimes \bar{N}\ma{I}_{\bar{N}^2}=(K/\bar{N})\ma{I}_{\bar{N}^2}$. Hence, it follows that $\ma{S}^{\text{H}}_{3}\ma{S}_{3}=\ma{I}_{\bar{N}^2Q}$.
Due to this column-orthogonality property, a simple matched filtering can replace the pseudo-inverse in step 3 of Algorithm 1 while offering optimum performance \cite{Clerckx_Arxiv_2024}.

\subsection{BD-RIS design for the BTALS algorithm}
The BTALS algorithm estimates the individual channel matrices from the 1-mode and 2-mode unfoldings $[\ten Y]_{(1)}$ and  $[\ten Y]_{(2)}$ according to (\ref{eq:ls_GH_sol}).

{\color{black} \emph{Property 2}: If no proportional columns exist in the set $\{\ten S^{(q)}_{..1}, \ldots, \ten S^{(q)}_{..K}\}$, $\forall q \in \{1, \ldots, Q\}$, meaning that the BD-RIS training tensor does not have proportional frontal slices, then $\ma{S}_1$ and $\ma{S}_2$ have full row-rank and are row-orthogonal, satisfying $\ma{S}^{\text{H}}_1\ma{S}_1=\ma{S}^{\text{H}}_2\ma{S}_2=K\ma{I}_{\bar{N}Q}$. 

To verify this property, first note that according to (\ref{eq:Zk1}) and (\ref{eq:blkdiag_Sk_basis}), the $k$-th frontal slice $\ten{S}^{(q)}_{..k}$ is orthogonal by construction, i.e., $\ten{S}^{(q)}_{..k}\ten{S}^{(q)\text{H}}_{..k}=\ten{S}^{(q)\text{H}}_{..k}\ten{S}^{(q)}_{..k}=\ma I_{\bar{N}}$, $k=1, \ldots, K$. Since $\ma{S}_1$ (resp. $\ma{S}_2$) concatenates the columns (resp. rows) of the $K$ slices $\{\ten S^{(q)}_{..k}\}$ (see (\ref{eq:Sq_unf1})-(\ref{eq:Sq_unf2})), they will have full rank $\bar{N}Q$ if their columns (resp. rows) are non-proportional, which implies $\ten S_{..k} \neq \alpha \ten S_{..k'}$, $\forall k \neq k'$, $k,k' \in \{1,\ldots, K\}$, where $\alpha$ is a scalar. 
The constraint $\ma{S}^{\text{H}}_1\ma{S}_1=\ma{S}^{\text{H}}_2\ma{S}_2=K\ma{I}_{\bar{N}Q}$ follows from the orthogonality of each matrix $\ten{S}^{(q)}_{..k}$, $k=1, \ldots, K$.}

To ensure Property 2 while preserving its structural properties, the BD-RIS training tensor is designed as
\begin{align}
\label{eq:tenS_q_rotation}
\ten{S}^{(q)}_{..k}= \mathrm{D}_k(\ma{W}^{(q)})\bar{\ten{S}}^{(q)}_{..k}\mathrm{D}_k(\ma{W}^{(q)*}),\, k=1, \ldots, K,
\end{align}
where $\bar{\ten{S}}^{(q)}_{..k}$ is given by (\ref{eq:blkdiag_Sk_basis}) and
$\ma W^{(q)}\doteq [\ma{w}^{(q)}_1,\ldots, \ma{w}^{(q)}_K]^{\text{T}} \in \mathbb{C}^{K \times \bar{N}}$ is a matrix of exponentials, whose $k$-th row is given by $\ma{w}^{(q)}_k\doteq [1 \,\, e^{j\psi^{(q)}_{k,1}}, \ldots, e^{j\psi^{(q)}_{k,\bar{N}-1}}]^{\text{T}}$, where $\{\psi^{(q)}_{k,n}\}$ are uniformly distributed between $[0, 2\pi)$, $k=1, \ldots, K$, $q=1,\ldots, Q$. The random phase shifts in (\ref{eq:tenS_q_rotation}) ensure that the $K$ frontal slices of the BD-RIS training tensor are non-proportional, regardless of the values of $\bar{N}$ and $Q$. Note also that the orthogonality of the frontal slices are preserved after these random phase shifts, i.e., $\ten{S}^{(q)}_{..k}\ten{S}^{(q)\text{H}}_{..k}=\ten{S}^{(q)\text{H}}_{..k}\ten{S}^{(q)}_{..k}=\ma I_{\bar{N}}$, $k=1, \ldots, K$.

\subsection{Discussion}
It is clear that the BTALS algorithm has less restrictive requirements on the training length $K$ than the BTKF algorithm. This comes from the fact that the latter only exploits the 3-mode unfolding of the received pilot tensor (as is the case with the LS method \cite{Clerckx_Arxiv_2024}), and channel decoupling is carried out after the LS filtering step, which is the bottleneck restricting the training length. On the other hand, the BTALS algorithm exploits the 1-mode and 2-mode unfoldings of the received pilot tensor to estimate the involved channel matrices directly. The associated LS steps involve more relaxed requirements for the training length $K$ compared to BTKF.

It is worth highlighting the trade-offs between BTKF and BTALS regarding computational complexity and processing delay. 
To calculate the complexity, we assume a complexity $\mathcal{O}(mn)$ for computing a rank-one approximation of a matrix $\ma{A} \in \bb{C}^{m \times n}$ \cite{flops_rank1}, 
while the Moore-Penrose pseudo-inverse of $\ma{A}$ has a cost $\mathcal{O}(m^2n)$. Hence, the BTKF algorithm requires $\mathcal{O}(Q \bar{N}^2M_RM_T)$ for computing the $Q$ rank-one approximations to find the individual channel estimates for each BD-RIS group, as shown in (\ref{prob:ls_kron_Q_rank_one}). In addition, since BTKF also includes a prior LS estimation step, its overall complexity corresponds to $\mathcal{O}(M_RTK \bar{N}^2Q)$. In contrast, the BTALS algorithm involves two LS estimation steps at each iteration, where each step requires the computation of two matrix inverses, as shown in steps 4 and 5 of Algorithm 2. Considering the dimensions of these matrix inverses and summing up their individual complexities, we arrive at a total cost of $\mathcal{O}(I_{\textrm{max}}  \bar{N}^2QK(M_R + M_T ))$, where $I_{\textrm{max}}$ denotes the maximum number of iterations assumed.

It is worth mentioning that the BTFK method accomplishes channel estimation in closed form utilizing $Q$ independent and parallel processing routines, which implies a shorter processing delay than BTALS. The latter consists of a sequential process of alternating estimation steps, where the channel matrices associated with the $Q$ BD-RIS groups are jointly estimated. Therefore, there is a longer processing delay than BTKF. From a practical perspective, optimized procedures for computing rank-one approximations and the possibility of parallel processing are attractive features of BTKF despite its higher training overhead than BTALS. Thus, there is a clear trade-off between both algorithms involving complexity, training overhead, and processing delay. 

 \section{Numerical results}\label{Sec:Numerical_Results}
We evaluate the performance of the proposed BTKF and BTALS receivers under different system setups. We also discuss the tradeoffs involving the proposed channel estimation methods and their superior performance compared to the  LS scheme \cite{Clerckx_Arxiv_2024}. We consider the \ac{NMSE} of the composite channel $ \ma{T} = \ma{H} \blockron \ma{G} \in \bb{C}^{M_RM_T \times \bar{N}^2Q}$. The \ac{NMSE} is given by 
\begin{align*}
    \text{NMSE}\big( \hat{\ma{T}} \big) = \frac{\fronorm{\ma{T} - \hat{\ma{T}} }^2 }{\fronorm{\ma{T}}^2}.
\end{align*}
As in \cite{Clerckx_Arxiv_2024}, we assume i.i.d. channels, and we compare the performance of the different methods for various parameter setups shown in each figure.

\begin{figure}[!t]
 \centering\includegraphics[scale=0.5]{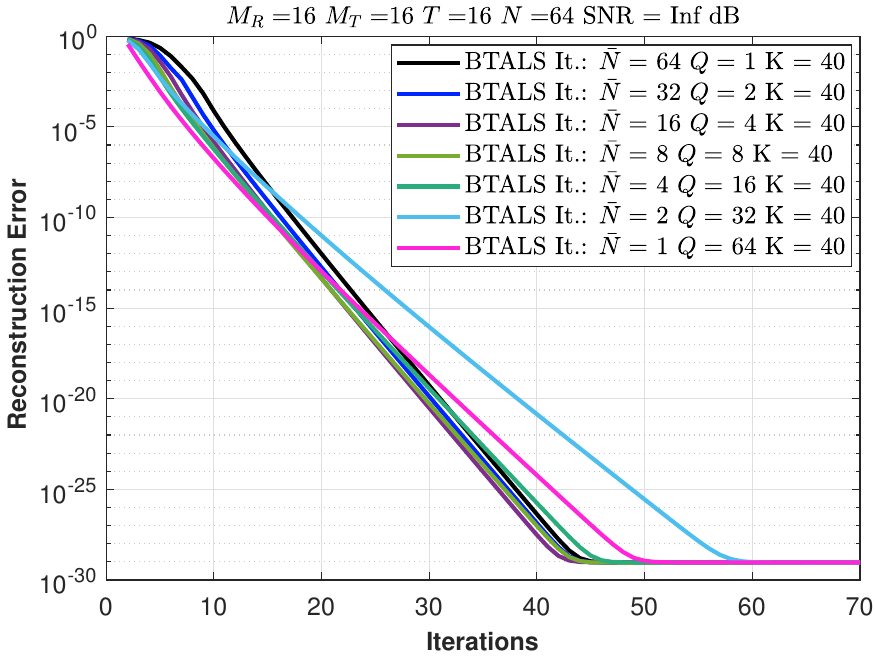}
  \caption{Reconstruction error $\epsilon$ per iteration for different setups.}
  \label{fig:BTALS_IT_recon}
  \end{figure}

\begin{figure}[!t]
  \centering
  \includegraphics[scale = 0.5]{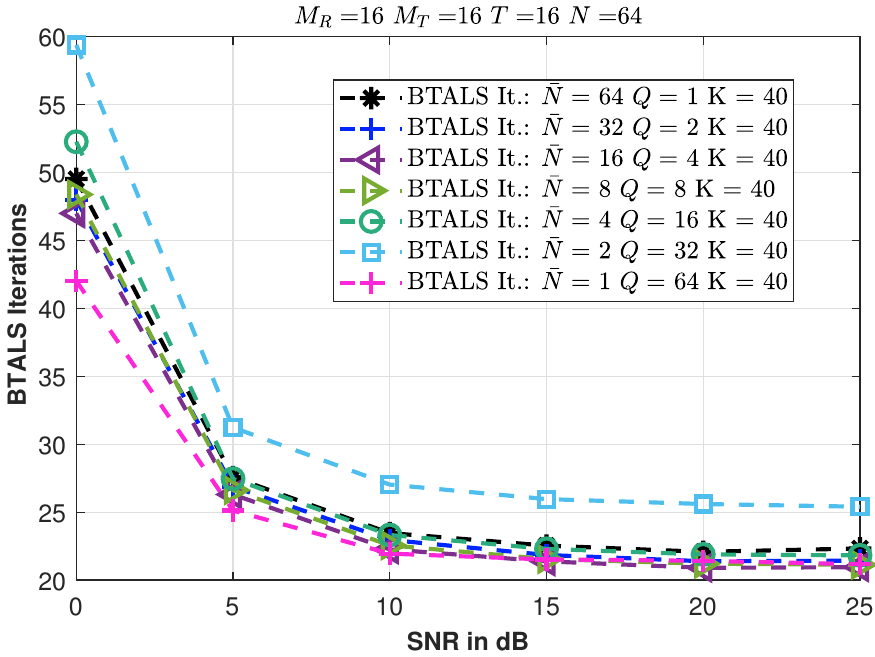}
  \captionof{figure}{Number of iterations to convergence \textit{vs.} SNR.}
  \label{fig:BTALS_vs_SNR_it}
\end{figure}

\begin{figure}[!t]
 \centering\includegraphics[scale=0.5]{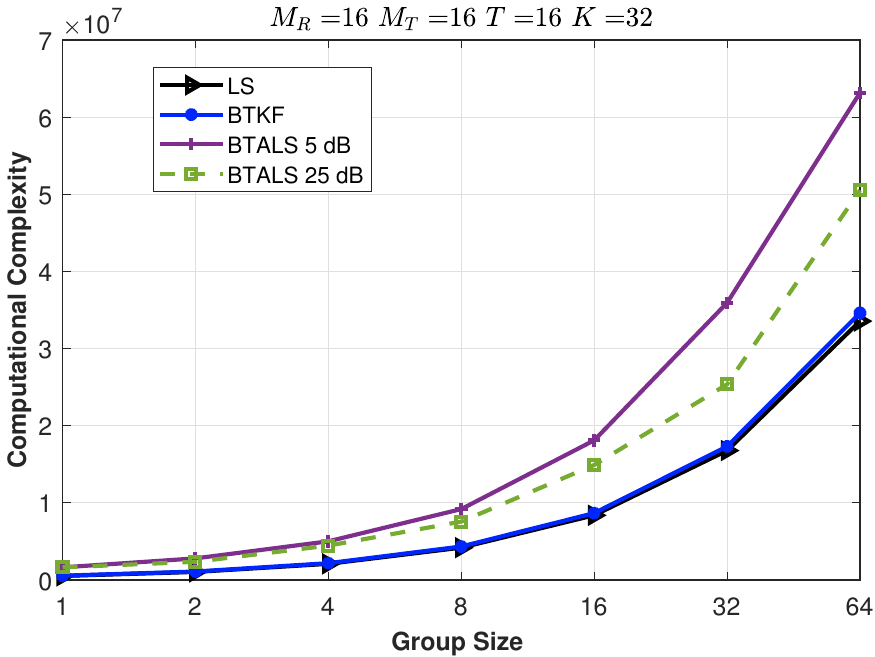}
	\caption{Computational complexity of LS \cite{Clerckx_Arxiv_2024}, BTKF, and BTALS for different group sizes $\bar{N}$.}
	\label{fig:CC_LS_BTKF_BTALS}
\end{figure}

\begin{figure}[!t]
 \centering\includegraphics[scale=0.5]{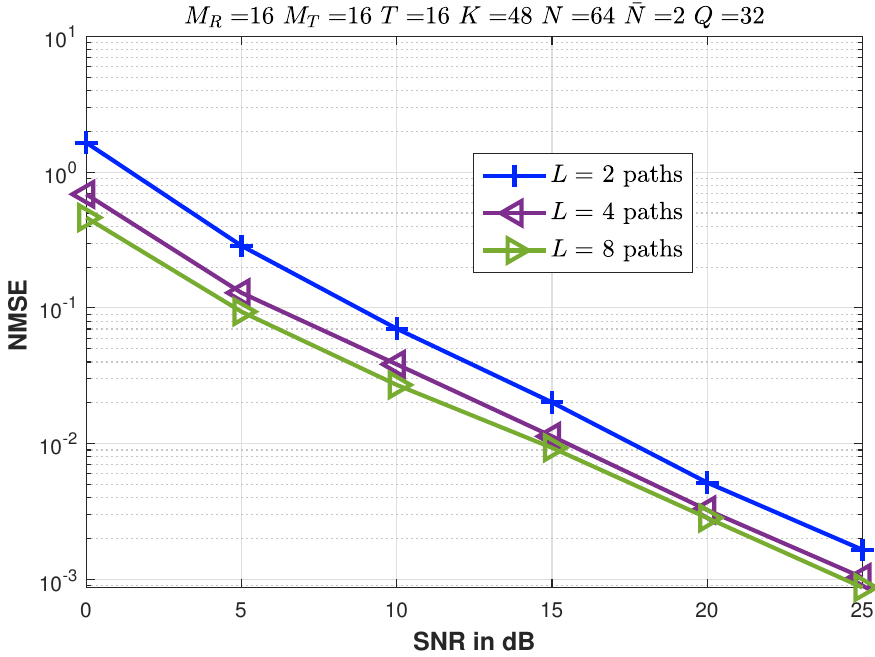}
	\caption{BTALS vs. SNR performance for low-rank channel $\ma{H}$, considering different number of channel paths.}
	\label{fig:BTALS_geo_chan}
\end{figure}

\begin{figure}[!t]
 \centering\includegraphics[scale=0.5]{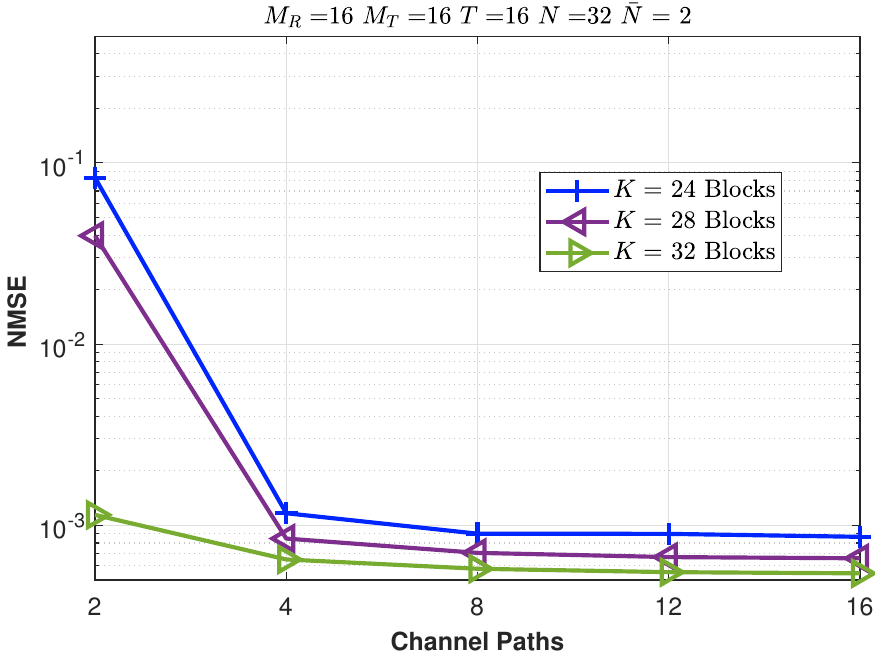}
	\caption{BTALS performance for low-rank channel $\ma{H}$ as a function of the number of channel paths and SNR of $25$ dB.}
	\label{fig:BTALS_geo_chan_vs_K}
\end{figure}

{\color{black}
In Figures \ref{fig:BTKF_vs_LS_fixed_K_N_64} to  \ref{fig:BTKF_vs_LS_fixed_K_N_64_vary_Mt}, we evaluate the performance of the proposed receiver BTKF with the baseline LS of \cite{Clerckx_Arxiv_2024}. It is worth mentioning that the BTALS is omitted here due to the fact that it achieves the same performance as the BTKF in the considered scenarios. Since the BTALS is the only receiver that can operate at a lower training overhead, we will discuss its performance in a separate section (Figure \ref{fig:BTALS_vs_K_snr_25dB} to \ref{fig:BTALS_geo_chan_vs_K}). For all results, orthogonal pilot sequences are assumed, and their length is fixed to its minimum value $T_{\text{min}} = M_T$.

\begin{figure}[!t]
	\centering\includegraphics[scale=0.5]{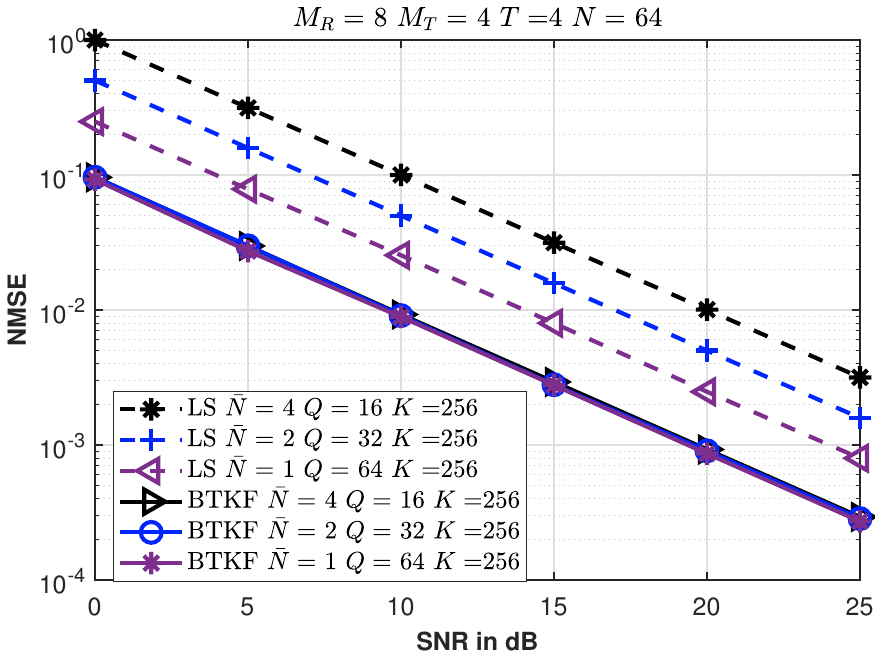}
	\caption{Comparison between BTKF and LS for different BD-RIS configurations.}
	\label{fig:BTKF_vs_LS_fixed_K_N_64}
\end{figure}

\begin{figure}[!t]
	\centering\includegraphics[scale=0.5]{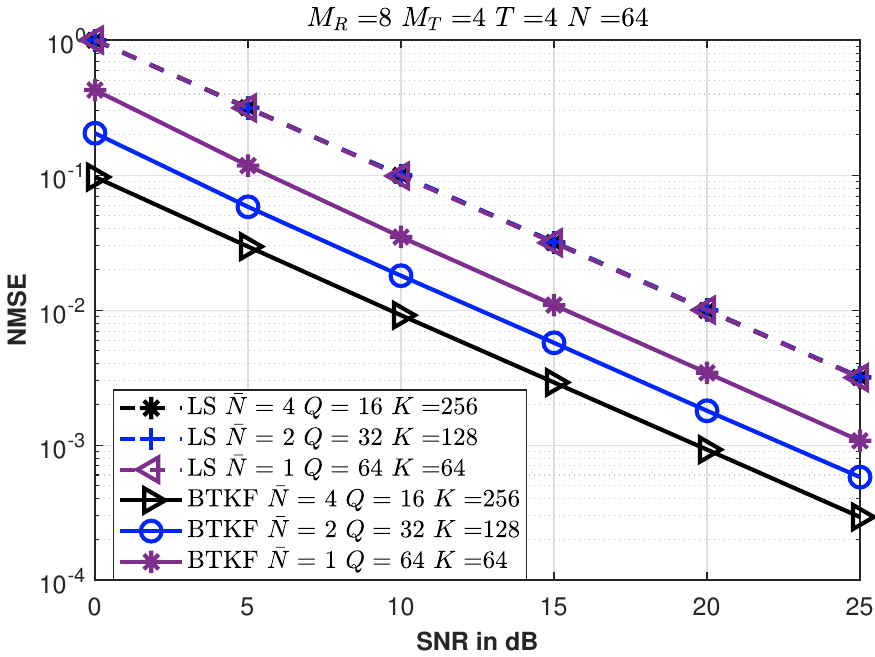}
	\caption{Comparison between BTKF and LS considering the minimum training overhead ($K_{\text{min}}$) for each configuration. }
	\label{fig:BTKF_vs_LS_vary_K_N_64}
\end{figure}

In Figure \ref{fig:BTKF_vs_LS_fixed_K_N_64}, we compare the NMSE of the composite channel by fixing $K= 256$, which is the minimum value ($K_{\text{min}} = \bar{N}^2Q$) for the configuration with largest group size $\bar{N}=4$. First, BTKF offers higher estimation accuracies than the baseline LS method, regardless of the group size. Indeed, conventional LS estimates the composite channel as a classical MIMO channel estimation problem, which is \emph{blind} to its inherent Kronecker structure. Thus, the performance of the reference method is sensitive to the number of groups and degrades as the group size ($\bar{N}$) increases, while the BTKF method is insensitive to it. Additionally, we can see that the performance gains of BTKF over LS increase with $\bar{N}$. Such a gain comes from the channel separation property of BTKF that exploits the Kronecker structure of the composite channel. 

\begin{figure}[!t]
	\centering\includegraphics[scale=0.5]{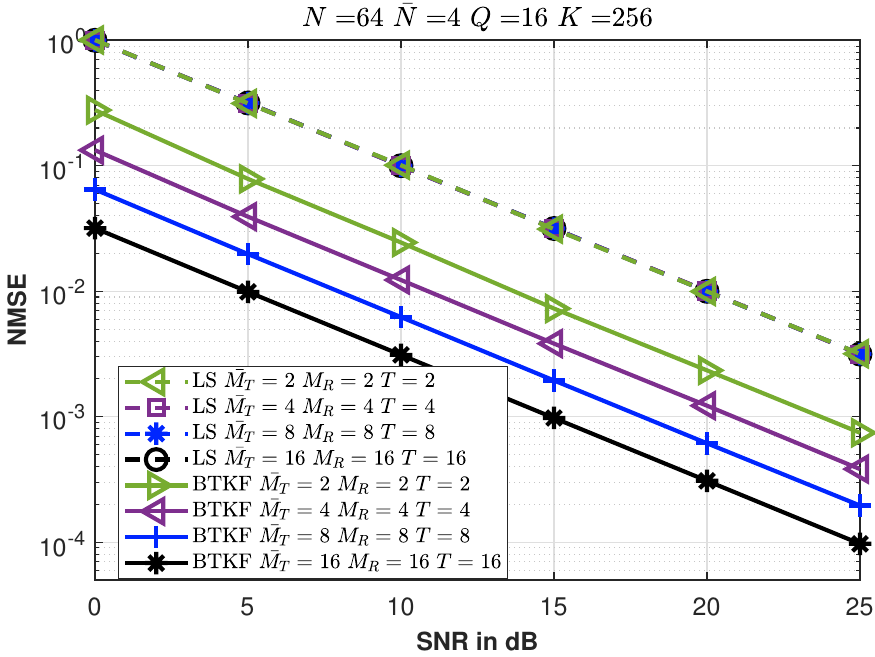}
	\caption{Comparison between BTKF and LS for different numbers of transmit and receive antennas.}
	\label{fig:BTKF_vs_LS_fixed_K_N_64_vary_Mt}
\end{figure}

\begin{figure}[!t]
 \centering\includegraphics[scale=0.5]{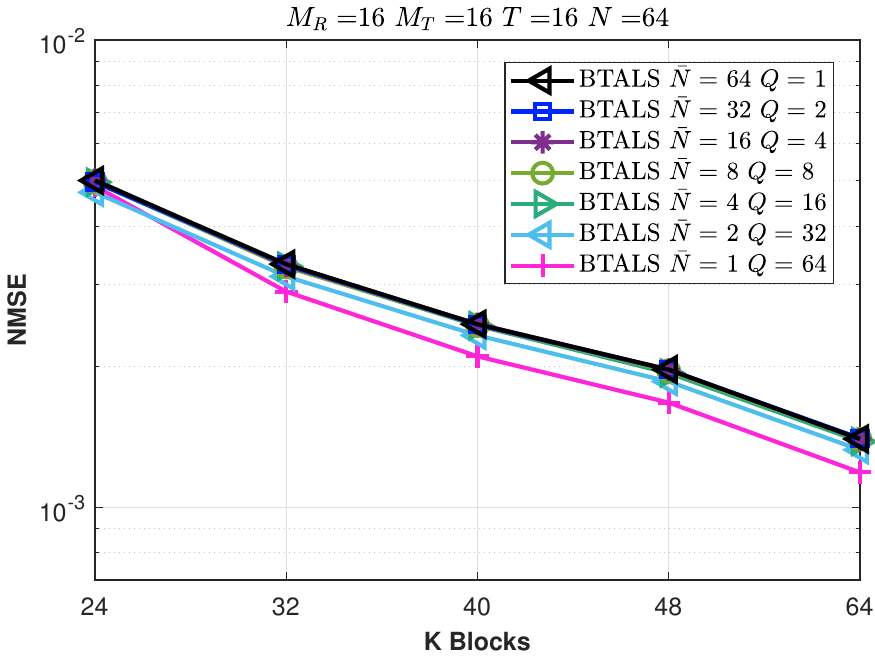}
	\caption{Performance of BTALS as a function of the number $K$ of training blocks. The SNR is fixed to 20 dB.}
	\label{fig:BTALS_vs_K_snr_25dB}
\end{figure}

In Figure \ref{fig:BTKF_vs_LS_vary_K_N_64}, we assume the minimum training overhead for each considered BD-RIS configuration, i.e., $K_{\text{min}}= \bar{N}^2Q$ for each setup that ensures a unique solution. It is interesting to note that the baseline LS scheme \cite{Clerckx_Arxiv_2024} has the same performance in all configurations. Hence, the estimated number of channel coefficients is the same in all the configurations considered in this figure. In contrast, the BTKF method estimates fewer channel coefficients than LS. The gap in the number of channel coefficients to be estimated increases as the group size increases, which explains the performance gap between the two methods as $\bar{N}$ increases. Note that, in the single-connected case ($\bar{N}=1$), the estimated channel coefficients are the same for LS and BTKF. However, the BTKF still offers a performance gain over LS, even in this case. From this set of results, one can conclude that as the group size increases, the noise rejection gain offered by the rank-one approximation step of BTKF also increases. In contrast, the LS method has nearly the same performance regardless of the group size since it does not take advantage of the Kronecker structure of the composite channel.

In Figure \ref{fig:BTKF_vs_LS_fixed_K_N_64_vary_Mt}, we fix the group size $\bar{N}=4$ and the total number of groups $Q=16$, implying $N=64$ RIS elements. In this experiment, we compare the performance of BTKF and LS for different numbers of antennas $M_T$ and $M_R$ at the transmitter and receiver, respectively. Note that the pilot sequence length is adjusted in each configuration to its minimum value ($T=M_T$). We observe that the performance of the LS estimator is the same regardless of the number of transmit/receive antennas. Indeed, although the pilot length increases to ensure orthogonality, the number of composite channel coefficients also increases with $M_T$ and $M_R$. This means the LS estimator cannot benefit from the added spatial degrees of freedom to improve channel estimation performance. In contrast, the BTKF method efficiently benefits from more transmit and receive spatial degrees of freedom by capitalizing on the Kronecker structure of the composite channel. Such a gain comes from the noise rejection gains provided by the rank-one approximation problems in (\ref{prob:ls_kron_Q_rank_one}), whose dimensions increase with $M_T$ and $M_R$, yielding more accurate estimates of $\ma{G}$ and $\ma{H}$ and, consequently, of the composite channel $\hat{\ma{T}}=\hat{\ma{H}} \blockron \hat{\ma{G}}$. 

\begin{figure}[!t]
	\centering\includegraphics[scale=0.5]{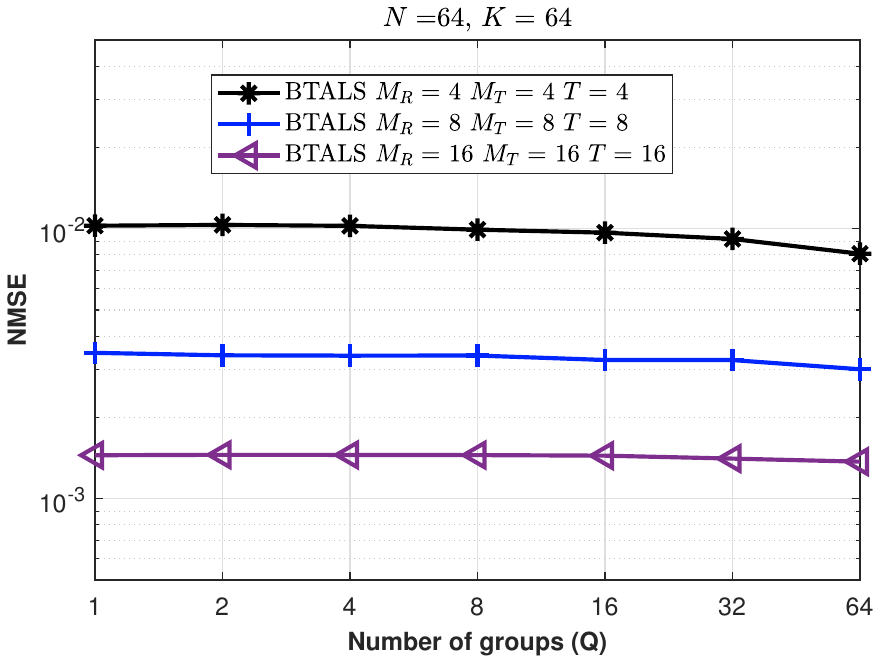}
	\caption{BTALS performance vs. the number $Q$ of groups.}
	\label{fig:BTALS_vary_Mt_Mr}
\end{figure}

\begin{figure}[!t]
\centering\includegraphics[scale=0.5]{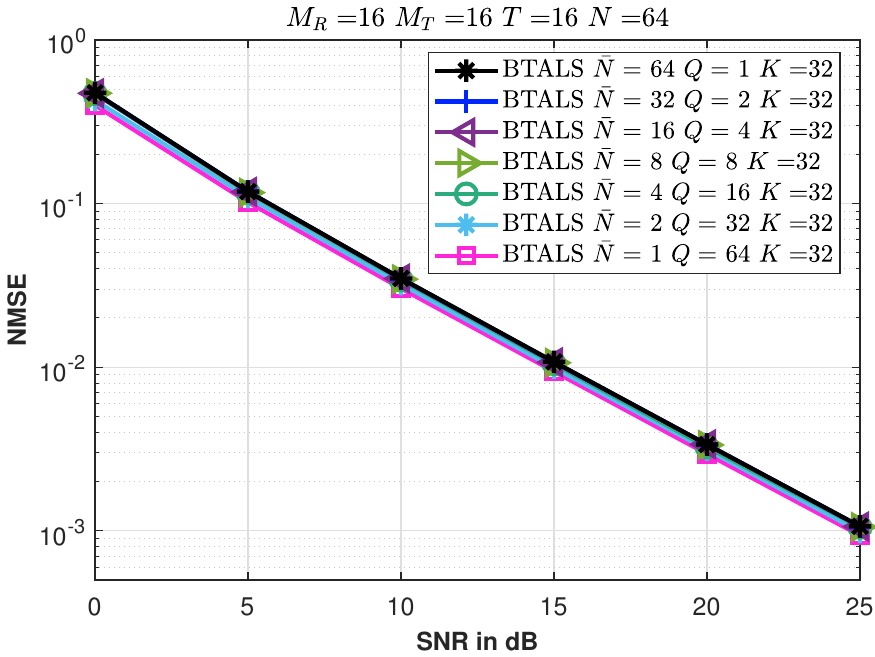}
	\caption{BTALS performance for different configurations.}
	\label{fig:BTALS_vary_Q_snr}
\end{figure}

In the following experiments (Figures \ref{fig:BTALS_vs_K_snr_25dB} to \ref{fig:BTALS_geo_chan_vs_K}), we turn our attention to the BTALS algorithm and study its performance for several system configurations. The focus is on low training overhead setups, where $K<<\bar{N}^2Q$. Hence, these figures do not show the results of the LS and BTKF methods since they cannot operate in the considered challenging setups over the full range of values of $K$ due to their more restrictive requirements on this training parameter.

Figure \ref{fig:BTALS_vs_K_snr_25dB} depicts the NMSE performance of BTALS as a function of the number $K$ of training blocks, by varying from $K=24$ to $K=128$, for a fixed SNR equal to $20$ dB. It can be observed that all configurations achieve very close performance. In particular, the performance gap between group-connected architectures and the fully connected architecture is negligible. The group-connected configurations have a slight performance gap ($\approx0.5$ dB) over the single-connected one ($\bar{N}=1$, $Q=64$). Note that for most of the system setups ($\bar{N}=\{4,8,16,32,64\}$), the range considered for the BD-RIS training length $K$ is far below the minimum value required by the LS and BTKF methods, which is $K_{\text{min}} = \bar{N}^2Q$ in each case. As an example, for the configuration $(\bar{N},Q)=(4,16)$, LS and BTKF would require $K=256$  blocks while for $(\bar{N},Q)=(64,1)$, they would require $K=4096$ blocks. These results corroborate the remarkable savings of training resources provided by the BTALS algorithm, which can operate over a broader set of choices for $K$ with significantly lower training overheads.
}

In Figure \ref{fig:BTALS_vary_Mt_Mr}, we show the performance of BTALS as a function of the number $Q$ of groups, going from the fully-connected case ($Q=1$) to the single-connected case ($Q=64$). We assume $N=\bar{N}Q=64$ RIS elements, $K = 64$ blocks, and an SNR of $20$ dB while considering configurations with different numbers of transmit and receive antennas. As expected, the performance increases as more transmit/receive antennas are used, showing the effectiveness of BTALS in converting spatial diversity gains at both link ends into more accurate channel estimates despite the increase in the number of estimated channel coefficients. These results corroborate the gains of tensor-based processing for BD-RIS channel estimation. We can also see that when the number of antennas and the transmit/receive increases, the gap among the single-, group-, and fully-connected architectures reduces. 


Figure \ref{fig:BTALS_vary_Q_snr} shows the NMSE performance of BTALS as a function of the SNR, assuming $N=64$ and different BD-RIS configurations (combinations of $\bar{N}$ and $Q$), including the single-connected architecture to the fully-connected one as extreme cases. In all configurations, the training length $K=32$ (corresponding to $N/2$) is much smaller than the product $\bar{N}^2Q$ that is the minimum overhead required by the LS and BTKF methods (both cannot operate in this case). We can see similar results in all the considered configurations. These results show that BTALS is an attractive solution in terms of training overhead compared to the competing schemes.

{\color{black}In Figures \ref{fig:BTALS_IT_recon} and \ref{fig:BTALS_vs_SNR_it}, we evaluate the convergence properties of the BTALS algorithm for different system setups. Figure \ref{fig:BTALS_IT_recon} corroborates the monotonic convergence of BTALS for all BD-RIS configurations, i.e., from the single-connected case ($\bar{N}=1$) to the fully-connected case ($\bar{N}= 64$), in the noiseless case. We can also observe that, regardless of the specific setup, the BTALS algorithm always converges to the machine's precision within 40-50 iterations. Figure \ref{fig:BTALS_vs_SNR_it} shows a different view of the convergence by depicting the required number of iterations to BTALS convergence for different system setups. As expected, the convergence is faster as the SNR increases. In particular, for a medium SNR above $10$ dB, convergence is always achieved with less than 30 iterations. It is worth mentioning that these results assume only $K=40$ blocks, which is significantly smaller than $\bar{N}^2Q$.}

In Figure \ref{fig:CC_LS_BTKF_BTALS}, we compare LS, BTKF, and BTALS in terms of computational complexity. We plot the complexity order of each algorithm as a function of the group size $\bar{N}$ for a BD-RIS with $N=64$ elements. Recall that $\bar{N}=1$ corresponds to the single-connected case, while $\bar{N}=64$ corresponds to the fully-connected case. As expected, the computational complexity increases with the group size in all cases. Moreover, since the convergence of BTALS is sensitive to the SNR (see Fig. \ref{fig:BTALS_vs_SNR_it}), its overall complexity will increase at low SNRs. This is not the case with BTKF, whose complexity does not depend on the SNR due to its closed-form nature. Such a complexity gap is the price BTALS pays to operate at a very low overhead compared to the BTFK and the LS methods. In addition, BTKF and LS have comparable complexities in this scenario since the complexity is dominated by the matched filtering step (the same for both LS and BTKF). 

{\color{black} Our proposed algorithms can also operate under spatially correlated channels or scattering-poor propagation (low-rank) channels. In Figure \ref{fig:BTALS_geo_chan}, we show the NMSE performance of} {\color{black}the proposed BTALS algorithm under multipath (geometric) channels. More specifically, we assume that the channel between the transmitter and the BD-RIS ($\ma{H}$) is modeled as
\begin{align*}
  \ma{H} = \sumx{l=1}{L}\ma{a}^{(\text{Tx})}_l \gamma_l\ma{a}^{(\text{RIS})\text{T}}_{l} \in \bb{C}^{M_T \times N},
\end{align*}
where $\ma{a}^{(\text{RIS})}_l \in \bb{C}^{N\times 1}$ and $\ma{a}^{(\text{Tx})}_l \in \bb{C}^{M_T \times 1}$ are the steering vectors associated with the arrival and departure angles at the BD-RIS and Tx, respectively. The angles are randomly generated between $[-\pi,\pi]$ according to a uniform distribution, while the $l$-th path amplitude $\gamma_l$ is a zero mean unit variance complex Gaussian random variable, $l=1,\ldots, L$. \textcolor{black}{The channel matrix $\ma{G}$ linking the BD-RIS to the receiver follows a  Rayleigh fading model, corresponding to a rich-scattering propagation.} {\color{black} Considering the rank-deficient channel case, improved performance is obtained when the number of channel paths linking the transmitter to the BD-RIS increases. This is expected since, in this case, the rank of $\ma{H}$ corresponds to the number $L$ of paths, where $L < \text{min}\{M_T,N\}$. The higher the $L$, the better the condition number of the channel matrix $\ma{H}$. Note, however, that the performance gap is not significant}. 

{\color{black} Figure \ref{fig:BTALS_geo_chan_vs_K} shows the performance of the BTALS as a function of the number of paths of $\ma{H}$ for different values of $K$ and an SNR of 25dB. The results show that poor scattering scenarios with few paths become more sensitive to the training length $K$. As the number of channel paths increases, the choice of this parameter has a negligible influence on performance.
}

{\color{black} The BTKF and BTALS algorithms have interesting tradeoffs. When training overhead is not critical, BTKF is preferable due to its lower complexity, closed-form solution, and small processing delay, since each rank-one factorization is independent and can be computed in parallel. On the other hand, when minimizing the training overhead is crucial (especially for strongly connected BD-RIS), BTALS is more attractive due to its flexibility in operating with considerably fewer training blocks than the LS and BTKF methods with no performance degradation.}

{\color{black}
\section{Conclusion and perspectives}\label{Sec:Conclusions}
The channel estimation problem for BD-RIS can be addressed from a tensor decomposition perspective. Recasting the received pilot signals as a block Tucker tensor model yields individual estimates of the involved channels by exploiting the multilinear structure of the received pilot signals. Decoupling the individual channels for BD-RIS has some key benefits. First, it improves CSI estimation accuracy over traditional LS estimation by capitalizing on the composite channel's inherent (block) Kronecker structure. The gains become more pronounced as the number of transmit and receive antennas increases. Second, the channel estimates are obtained with significantly lower training overheads than the LS method. {\color{black} Perspectives include studying channel estimation methods in the presence of mutual coupling, which complicates the channel model and thus requires new BD-RIS designs to facilitate the training process. In this context, mutual coupling-aware channel training and performance optimization of the proposed channel estimation algorithms deserve further study. The applicability of the proposed approach to other BD-RIS architectures, beyond group-connected ones, is also interesting to explore.}
{\color{black}  Also, the development of parametric channel estimation algorithms that explicitly exploit the geometrical structure of the involved channel matrices is of interest.} {\color{black} Finally, the study of identifiability for block Tucker decomposition with a known (and structured) core tensor is still an open problem, and the derivation of necessary and sufficient conditions related to the minimum value of $K$ ensuring unique channel estimates deserves further research.}
}

}

\renewcommand\baselinestretch{.87}

\bibliographystyle{IEEEtran}
\bibliography{ref.bib}

\end{document}